\documentclass[fleqn,usenatbib]{mnras}

\usepackage{newtxtext,newtxmath}

\usepackage[T1]{fontenc}

\DeclareRobustCommand{\VAN}[3]{#2}
\let\VANthebibliography\thebibliography
\def\thebibliography{\DeclareRobustCommand{\VAN}[3]{##3}\VANthebibliography}

\DeclareRobustCommand{\ion}[2]{%
\relax\ifmmode
\ifx\testbx\f@series
{\mathbf{#1\,\mathsc{#2}}}\else
{\mathrm{#1\,\mathsc{#2}}}\fi
\else\textup{#1\,{\mdseries\textsc{#2}}}%
\fi}

\usepackage{graphicx}	
\usepackage{amsmath}	
\usepackage{CJKutf8}
\newcommand{\chinese}[1]{\begin{CJK}{UTF8}{gbsn}\normalfont #1\end{CJK}} 

\def \update {\textcolor{black}}
\def \updateagain {\textcolor{black}}

\title[Spatially-resolved environmental quenching]{The SAMI Galaxy Survey: Using concentrated star-formation and stellar population ages to understand environmental quenching}

\author[D. Wang et al.]{Di Wang (\chinese{王迪}),$^{1,2}$\thanks{E-mail: di.wang@sydney.edu.au}
Scott M. Croom,$^{1,2}$
Julia J. Bryant,$^{1,2,3}$
Sam P. Vaughan,$^{1,2}$
Adam L. Schaefer,$^{4}$
\newauthor
Francesco D'Eugenio,$^{5}$
Stefania Barsanti,$^{6,2}$
Sarah Brough,$^{7,2}$
Claudia del P. Lagos,$^{8,2}$
Anne M. Medling,$^{9,2}$
\newauthor
Sree Oh,$^{6,2}$
Jesse van de Sande,$^{1,2}$
Giulia Santucci,$^{7,2}$
Joss Bland-Hawthorn,$^{1,2}$
Michael Goodwin,$^{11}$
\newauthor
Brent Groves,$^{8,2}$
Jon Lawrence,$^{11}$
Matt S. Owers,$^{2,10,12}$
Samuel Richards$^{1}$
\\
\\
$^{1}$Sydney Institute for Astronomy (SIfA), School of Physics, The University of Sydney, NSW, 2006, Australia\\
$^{2}$ARC Centre of Excellence for All Sky Astrophysics in 3 Dimensions (ASTRO 3D)\\
$^{3}$Australian Astronomical Optics, Astralis-USydney, School of Physics, University of Sydney, NSW 2006, Australia\\
$^{4}$Max-Planck-Institut f{\"u}r Astrophysik, Karl-Schwarzschild-Str. 1, D-85748 Garching, Germany\\
$^{5}$Sterrenkundig Observatorium, Universiteit Gent, Krijgslaan 281 S9, B-9000 Gent, Belgium\\
$^{6}$Research School of Astronomy and Astrophysics, Australian National University, Canberra, ACT 2611, Australia\\
$^{7}$School of Physics, University of New South Wales, NSW 2052, Australia\\
$^{8}$ICRAR, The University of Western Australia, Crawley WA 6009, Australia\\
$^{9}$Ritter Astrophysical Research Center University of Toledo Toledo, OH 43606, USA\\
$^{10}$Department of Physics and Astronomy, Macquarie University, NSW 2109, Australia\\
$^{11}$Australian Astronomical Optics - Macquarie, Macquarie University, NSW 2109, Australia\\
$^{12}$Astronomy, Astrophysics and Astrophotonics Research Centre, Macquarie University, Sydney, NSW 2109, Australia\\
}

\date{Accepted XXX. Received YYY; in original form ZZZ}

\pubyear{2022}

\begin{document}
\label{firstpage}
\pagerange{\pageref{firstpage}--\pageref{lastpage}}
\maketitle

\begin{abstract}
We study environmental quenching using the spatial distribution of current star-formation and stellar population ages with the full SAMI Galaxy Survey. By using a star-formation concentration index [$C$-index, defined as $\log_{10}(r_\mathrm{50,H\alpha}/r_\mathrm{50,cont})$], we separate our sample into regular galaxies ($C$-index$\geq-$0.2) and galaxies with centrally concentrated star-formation (SF-concentrated; $C$-index<$-0.2$). Concentrated star-formation is a potential indicator of galaxies currently undergoing `outside-in' quenching. Our environments cover ungrouped galaxies, low-mass groups ($M_{200}\leq10^{12.5} M_{\odot}$), high-mass groups ($M_{200}$ in the range $10^{12.5-14} M_{\odot}$) and clusters ($M_{200}$>$10^{14} M_{\odot}$). We find the fraction of SF-concentrated galaxies increases as halo mass increases with 9$\pm$2 per cent, 8$\pm$3 per cent, 19$\pm$4 per cent and 29$\pm$4 per cent for ungrouped galaxies, low-mass groups, high-mass groups and clusters, respectively. We interpret these results as evidence for `outside-in' quenching in groups and clusters. To investigate the quenching time-scale in SF-concentrated galaxies, we calculate light-weighted age ($Age_\mathrm{L}$) and mass-weighted age ($Age_\mathrm{M}$) using full spectral fitting, as well as the $D_\mathrm{n}$4000 and $H\delta_\mathrm{A}$ indices. We assume that the average galaxy age radial profile before entering a group or cluster is similar to ungrouped regular galaxies. \update{At large radius (1-2 $R_\mathrm{e}$), SF-concentrated galaxies in high-mass groups have older ages than ungrouped regular galaxies with an age difference of 1.83$\pm$0.38 Gyr for $Age_\mathrm{L}$ and 1.34$\pm$0.56 Gyr for $Age_\mathrm{M}$}. This suggests that while `outside-in' quenching can be effective in groups, the process will not quickly quench the entire galaxy. In contrast, the ages at 1-2 $R_\mathrm{e}$ of cluster SF-concentrated galaxies and ungrouped regular galaxies are consistent (\update{difference of} $0.19\pm$0.21 Gyr for $Age_\mathrm{L}$, 0.40$\pm$0.61 Gyr for $Age_\mathrm{M}$), suggesting the quenching process must be rapid. 

\end{abstract}




\section{Introduction}
Many studies have shown that both galaxy stellar mass and the local environmental density in which a galaxy resides affect its growth, star-formation quenching, and morphology \citep[e.g.][]{Peng_2010,Peng_2012}. Environment influences gas accretion, gas removal and galaxy interactions, which then affect galaxy morphology \citep[e.g.][]{Dressler1980}, current star-formation \citep[e.g.][]{Lewis_2002} and star-formation history \citep[SFH; e.g.][]{Aird_2012}. In particular, we see that star-formation is significantly suppressed in higher environmental densities (groups and clusters; e.g. \citealp{Davies2019}) compared to galaxies in ungrouped regions. However, the detailed physical mechanisms responsible for the environmental suppression of star-formation remain unclear. 

Ram-pressure stripping (RPS) is a quenching process in which the interstellar medium (ISM) of a galaxy has a kinetic interaction with the intra-cluster medium (ICM) and forces the gas out of the galaxy \citep[e.g.][]{Gunn1972}. Studies of the impact of the environment on star-formation often focus on galaxy clusters where RPS is more clearly seen. For example, some Virgo cluster studies \citep[e.g.][]{Koopmann_2004, Koopmann_2006} have shown that the reduction of total star-formation in galaxies is caused by gas in discs stripped by the ICM. More recently, \citet{Chung2017} showed Virgo cluster star-forming (SF) galaxies being depleted of cold gas by RPS. Additionally, large-scale single-fibre spectroscopic galaxy surveys (e.g. the Sloan Digital Sky Survey; SDSS; \citealp{York_2000,Abazajian_2009}) point to environmental quenching being driven by processes that primarily act on satellite galaxies in haloes \citep[e.g.][]{Balogh2000,Ellingson2001,Peng_2012, Wetzel_2013}. 

While the influence of the environment has been demonstrated in galaxy clusters, environmentally driven star-formation quenching also occurs in less dense groups and pairs \citep[e.g.][]{Barsanti2018, Davies2019, Dzudzar2021}. For example, \citet{Lewis_2002} with the 2dF Galaxy Redshift Survey showed that environmental influences on galaxy properties are not restricted to cluster cores, but are effective in all groups where the projected densities exceed $\sim$1 galaxy Mpc$^{-2}$. \citet{Barsanti2018} observed that SF galaxies in groups have a smaller specific star-formation rate (sSFR) than ungrouped galaxies. \citet{Vazquez_2020} found that star-formation quenching is a significant and ongoing process as galaxies fall into galaxy groups. \citet[]{Oh_2018} found that early-type galaxies which form the majority population in galaxy clusters, have started or even finished star-formation quenching before they fall into clusters, implying there is preprocessing star-formation quenching happened in less dense environments before galaxies fall into clusters. For pairs, star-formation can be enhanced through close interactions \citep[e.g.][]{Woods_2010, Patton2013, Li2008, Davies2016}. \citet{Scudder2012} used SDSS pairs of galaxies to find that \update{star-formation} is enhanced in major merger systems, hinting that the pair mass ratio is significant in the modification of SFH through galaxy interactions. 

Optical integral field spectroscopy (IFS) is a crucial tool in understanding spatially-resolved star-formation. Early work by \citet{Brough_2013} used 18 galaxies in the Galaxy And Mass Assembly (GAMA; \citealt{Driver2011}) regions observed by the SPIRAL integral field unit (IFU) on the Anglo-Australian Telescope (AAT) without finding a relationship between the total star-formation rate (SFR) and environment. This observation would imply that any mechanism that transforms galaxies in dense environments must be rapid or have happened a long time ago in the local universe. However, using larger samples from the Sydney-Australian Astronomical Observatory Multi-object Integral Field Spectrograph (SAMI) Galaxy Survey, \citet[]{Medling2018} found that galaxies in denser environments show decreased sSFR in their outer regions consistent with environmental quenching. \citet{schaefer2019} explored the connection between star-formation and environment by using a larger sample of SAMI group galaxies in the GAMA regions. They found that in high-mass groups, SF galaxies with stellar mass $M_{\star} \sim 10^{10} M_{\odot}$ have centrally concentrated star-formation which suggests that they may be undergoing environmental quenching. \citet{Owers2019} showed cluster galaxies have indications of `outside-in' quenching by RPS within 8 SAMI clusters. The SDSS-IV Mapping Nearby Galaxies at Apache Point Observatory (MaNGA) survey has also been used to investigate star-formation quenching \citep[e.g.][]{Goddard2017, Ellison2018}. \citet{Bluck_2020} concluded that both intrinsic and environmental quenching must incorporate significant starvation of the gas supply.

Apart from SF, many galaxies have an active galactic nuclei (AGN) or low ionization nuclear emission-line regions (LINERs). This is important for the analysis of star-formation, as AGN in massive galaxies have been proposed as quenching agents by either expelling or heating the gas. The central AGN emission can also mask the flux from star-formation, or there may be no central star-formation, just AGN. In many cases, several sources of ionization will contribute to a single spectrum. If we simply exclude spaxels with AGN-like emission, the star-formation will be underestimated. Therefore, it is important to correct the flux due to AGN to capture the correct star-formation distribution. SF galaxies, AGNs and LINERs form separate branches on the Baldwin, Phillips $\And$ Terlevich (BPT) diagram \citep[ionisation diagnostic diagram;][]{Baldwin1981BPT,Kauffmann2003,Kewley_2001}. Recent studies have decomposed the emission-lines in galaxies hosting AGNs, to estimate the fraction of flux from SF and AGN-excited components in each spectrum \citep[e.g.][]{Davies2016}. \citet{Belfiore2016} used the [S\textsc{ii}]/H$\alpha$ ratio to quantify the residual star-formation in LINER-like regions. They define the typical line ratios for SF and LINER emission and compare these to the measured line ratios to estimate the fraction of star-formation in each spectrum. In this paper, we will use similar approaches to correct for LINER or AGN-like emissions.

However, past IFS studies have focused on either field/groups or field vs cluster environments: until now, there is no consistent study of resolved star-formation across \textit{all environments} (from field, groups to clusters). We aim to address this problem by leveraging the comprehensive environment coverage of the SAMI Galaxy Survey. In addition, we will combine SFRs with stellar-population properties, thus combining the `instantaneous' approach to quenching with a time-integrated analysis.

In this work, we study how spatially-resolved star-formation quenching depends on the environment by using a star-formation concentration index as a function of environment (including cluster, group and ungrouped) using SAMI IFU data. The primary goal is to use spatially-resolved H$\alpha$ emission as a star-formation distribution indicator to probe the star-formation quenching relationship with different environmental densities (including halo masses and the fifth-nearest-neighbour surface densities). Our second goal is to use stellar population measurement (ages from full spectral fitting and $D_\mathrm{n}$4000, $H\delta_\mathrm{A}$ indices) radial profiles to find quenching time-scales in different halo mass intervals. The paper is arranged as follows: we describe SAMI data and sample selection in Section~\ref{sec:dataandsampleselection}, SF properties and stellar population measurements in Section~\ref{sec:star_forming}. In Sections~\ref{sec:result_spatialSF} and \ref{sec:result_quenching}, we discuss the main outcomes of our research. In Section~\ref{sec:discussion}, we discuss our findings. In Section~\ref{sec:conclusion}, we summarize our conclusions. Throughout this work we assume $\Omega_\mathrm{m}$=0.3, $\Omega_\mathrm{\Lambda}$=0.7 and $H_\mathrm{0}$=70 $\mathrm{km s^{-1} Mpc^{-1}}$ as cosmological parameters.

\section{Data and Sample Selection} \label{sec:dataandsampleselection}
\subsection{SAMI Galaxy Survey}
The SAMI Galaxy Survey is an IFS project using the 3.9-m AAT. SAMI \citep[][]{Croom2012} has a 1-degree diameter field of view using 13 optical fibre bundles (hexabundles; \citealt{Bland2011,Bryant2011,Bryant2014}). Each bundle combines 61 optical fibres that cover a circular field of view with a 15$''$ diameter on the sky. These optical fibres feed into the AAOmega spectrograph \citep[][]{Sharp2006}. The SAMI Galaxy Survey observations took place between 2013 and 2018. The raw telescope data is reduced into two cubes using the 2dfDR pipeline \citep[]{AST2015}, together with a purposely written python pipeline \citep[][]{Allen2014} for the later stages of reduction \citep[][]{Sharp2015}. The blue cubes cover a wavelength range of 3700$-$5700 \AA $ $ with a spectral resolution of R = 1812 ($\sigma$ = 70 km \  s$^{-1}$), and the red cubes cover a wavelength range of 6250$-$7350 \AA $ $ with a spectral resolution of R = 4263 ($\sigma$ = 30 km \ s$^{-1}$) at their respective central wavelengths \citep[][]{Sande_2017}. In this paper, we use the third and final data release (DR3; \citealp{Croom_2021}) of all SAMI observations, together with value-added products such as emission-line fits, stellar population measurements and stellar kinematics. 

The SAMI Galaxy Survey sample is selected from the GAMA survey \citep[]{Driver2011} regions and 8 cluster regions. The sample input catalogue is built from the GAMA Survey in three equatorial regions \citep[]{Bryant2015}. The 8 massive clusters are chosen from the 2 Degree Field Galaxy Redshift Survey \citep[]{Colless2001} as well as the Sloan Digital Sky Survey (\citealt{York_2000}; \citealt{Abazajian_2009}) and are described by \citet{Owers2017}. The selection of specific clusters increases the dynamic range of galaxy environment probed. SAMI galaxies were targeted based on cuts in the redshift-stellar mass plane \citep[]{Bryant2015}. The primary sample is limited to redshift $z$ < 0.095 in which observations reached high and uniform completeness. A secondary sample included high-mass galaxies and higher redshift ($z$ < 0.115) in which galaxies were observed at a lower priority. In total, the sample contains 2100 observed GAMA galaxies (1940 primary, 160 secondary) and 888 observed cluster galaxies (756 primary, 132 secondary). There are also 80 galaxies as fill-in galaxies observed last which are mostly pairs. The observations reach 80.6 per cent completeness in the GAMA region and 87 per cent completeness in the cluster region for good galaxy targets from the SAMI input catalogue. The observed galaxies are representative of the input catalogue in redshift, stellar mass, effective radius ($R_\mathrm{e}$; major axis in arcsec) and 5th nearest neighbour density $\Sigma_{5}$/Mpc$^{2}$ as demonstrated by \citet[]{Croom_2021}.

\subsection{Emission-line fitting}
SAMI DR3 data products include cubes, binned cubes and aperture spectra \citetext{\citealp{Croom_2021}; earlier data releases are described in \citealp{Allen2015}, \citealp{Green2018}, \citealp{Scott2018}}. To produce the data products, the spectral continuum is fit using the Penalized PiXel-Fitting code (pPXF; \citealp{cappellari2007,Cappellari2017}) and the MILES single stellar population (SSP) spectral library \citep[][]{Vazdekis2010} on Voronoi-binned data \citep[the algorithm {\sc vorbin} from][]{Cappellari2017}. Then the full-resolution cube is refitted using a limited set of templates with priors on the weights as described in \citet{Owers2019}. The SAMI data cubes include seven strong emission-lines. By using version 1.1 of the LZIFU software package \citep[][]{Ho2016}, [O\textsc{ii}]3726+3729, H${\beta}$, [O\textsc{iii}]5007, [O\textsc{i}]6300, H${\alpha}$, [N\textsc{ii}]6583,  [S\textsc{ii}]6716 and [S\textsc{ii}]6731 are fitted in each unique spatial element with one to three Gaussian component profiles. All emission-lines are fitted simultaneously with their relative strengths and consistent velocities and velocity dispersions. A trained Neural Network LZCOMP \citep[][]{Hampton2017} then determines which one to three Gaussian components are necessary to describe the observed emission-line structure. For the H$\alpha$ line which is in the higher spectral resolution red spectrograph, SAMI DR3 provides multi-Gaussian fits for the decomposed flux of each component. Flux from multi-component fits are not necessary contributed only by SF and the components are not separately relatable to physical processes. Therefore, we use 1-component LZIFU fits in our study to capture good estimates of total flux. In addition to the emission-line products, SAMI DR3 also provides extinction maps derived from the Balmer decrement, classification maps and star-formation maps \citep{Medling2018}. 

\begin{figure}
	\includegraphics[width=\columnwidth]{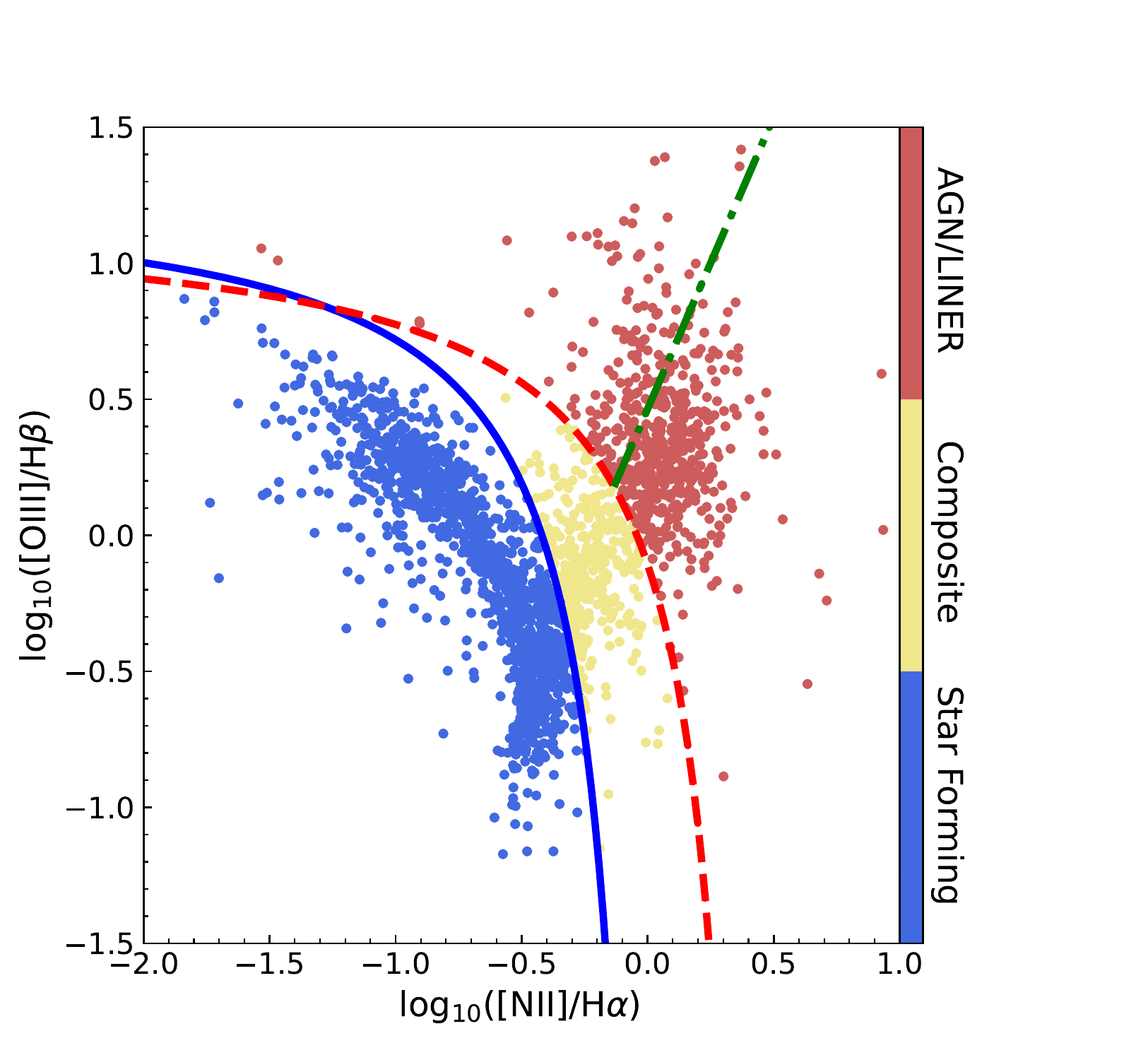}
    \caption{Ionization diagnostic diagram \citep[BPT;][]{Baldwin1981BPT} where each point shows the emission-line measurements derived from the central spaxel using a 1-component fit from the SAMI sample. We only include the central spaxels of each galaxy which have all emission-lines S/N > 2. The red dashed curve is the theoretical maximal SF line in \citet{Kewley_2001}. The solid blue curve is from \citet[]{Kauffmann2003}. The green dot-dashed line is separates \update{Seyfert galaxies} from LINERs \citep[][]{Kewley2006}. We use the selection boundaries to separate galaxies into SF (blue points), composite (yellow points) and LINER/AGN (red points).  }\label{fig:centre_pixcel_bpt}
\end{figure}

\begin{figure}
    \includegraphics[width=\columnwidth]{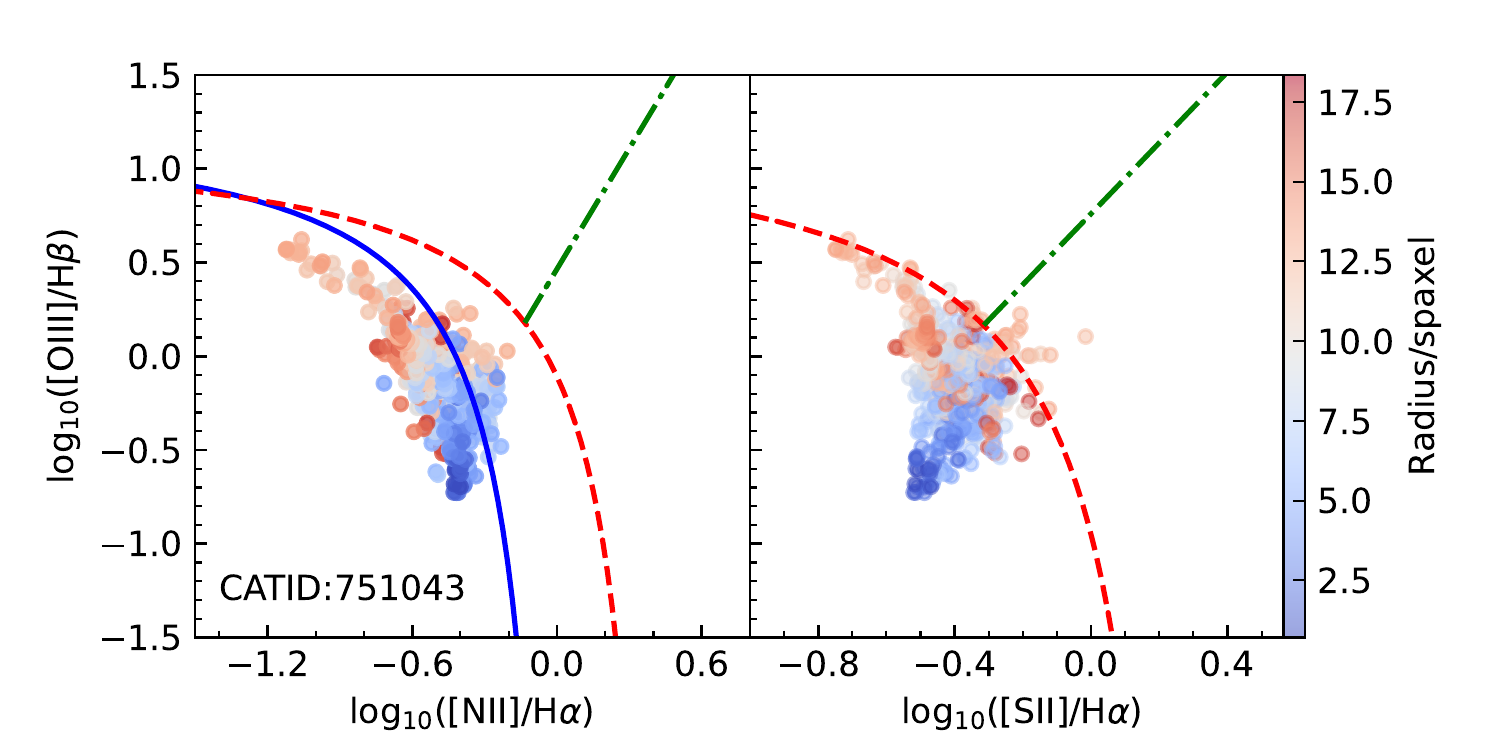}\\
    \includegraphics[width=\columnwidth]{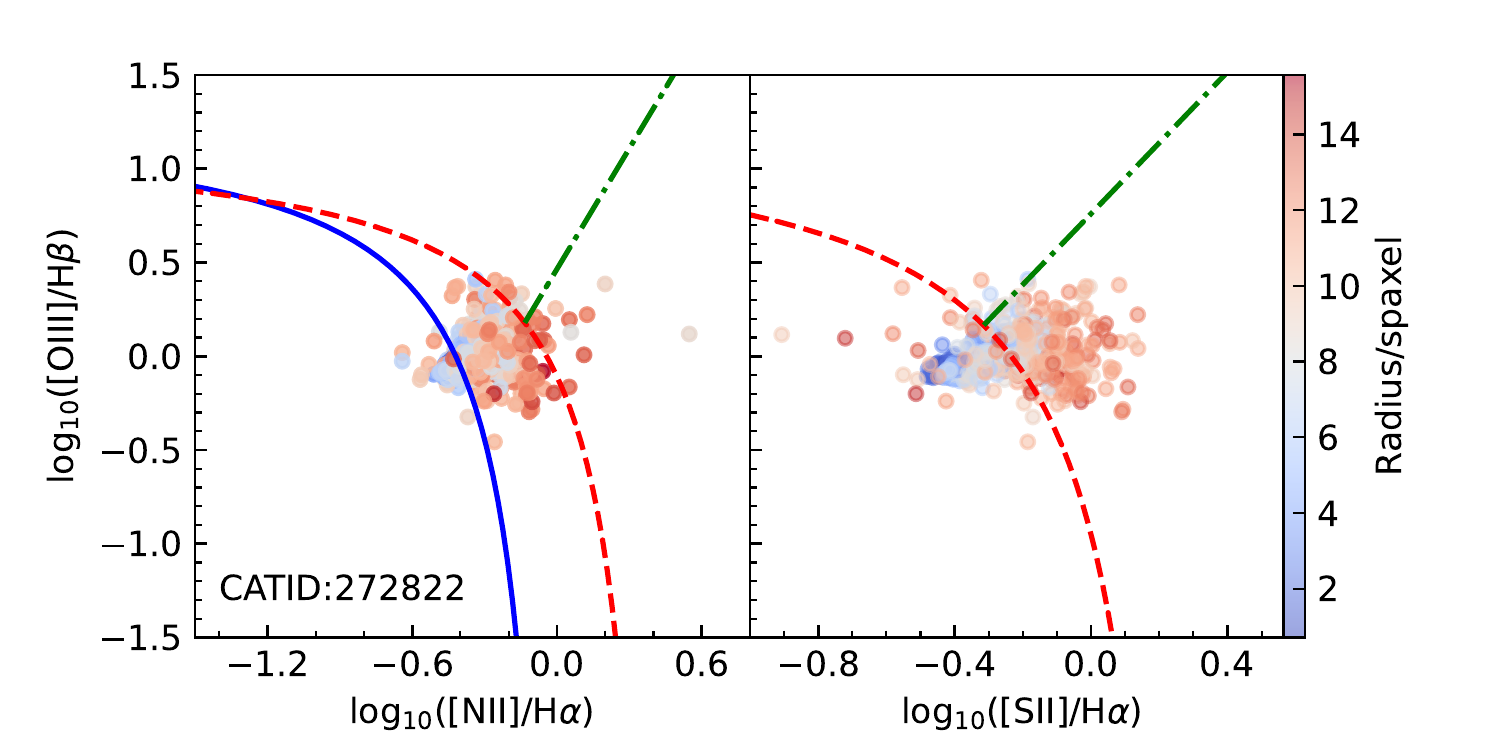}
    \caption{Example \update{of} spatially-resolved ionization diagram: Top panel: Galaxy 751043 is in \update{the} p$\_$SF sub-sample. Bottom panel: Galaxy 272822 is in the c$\_$SF sub-sample. The left panels show [O\textsc{iii}]/H$\beta$ versus [N\textsc{ii}]/H$\alpha$ emission-line ratios and the right panels show [O\textsc{iii}]/H$\beta$ versus [S\textsc{ii}]/H$\alpha$ emission-line ratios in the spaxels of each galaxy which have all emission-line S/N > 2. The points are colour-coded by radius in the galaxy. The blue, red, and green lines are the same as in Fig.~\ref{fig:centre_pixcel_bpt}\update{.}}\label{fig:SF_bpt}
\end{figure}

\begin{table*}
	\centering
	\caption{The result of galaxy selection including the number of galaxies in each sub-sample. The full sample (Full\_sam) is the main sample used throughout this paper. The Full\_SF sample is separated into pure SF (p$\_$SF) and central SF (c$\_$SF). We apply our AGN/LINER correction to all c$\_$SF and c$\_$L/A galaxies in this work.}
	\label{tab:sampletable}
	\begin{tabular}{cccc} 
		\hline
		Name & Abbreviation & Description & Numbers \\
		
		\hline
		Full Catalogue & Full$\_$ cat & Full SAMI sample including the GAMA and Cluster regions & 2988 \\
		Full sample & Full$\_$sam & All galaxies with $\mathrm{EW}_\mathrm{H\alpha}$ ${\ge}$ 1 \AA $ $ and $R_\mathrm{e}$ < 15 $''$, ellipticity < 0.7,  & 778\\
		$ $& $ $& seeing/$R_\mathrm{e}$ < 0.75, seeing < 4 $''$, \update{log(sSFR/yr$^{-1}$) > $-11.25$, not pure AGN } & $ $\\
		Full Star-forming & Full\_SF & SF galaxies in Full$\_$sam (central spaxel beneath the Kauffmann line)  & 719 \\
		Pure Star-forming & p$\_$SF &  Pure SF galaxies in \update{Full\_SF} & 653 \\
		Central Star-forming & c$\_$SF &  Central SF with LINER/AGN on the outskirts in \update{Full\_SF} & 66 \\
		Central LINER/AGN & c$\_$L/A & Central LINER/AGN with extended \update{star-formation} in Full$\_$sam & 59 \\
		\hline
	\end{tabular}
\end{table*}

\subsection{Sample Selection}
\label{sec:sample_section}
Our sample galaxies are from the full SAMI Galaxy Survey. These galaxies comprise SAMI data release three (DR3) primary and secondary samples. There are 2100 galaxies in the GAMA region and 888 galaxies in the cluster regions. \update{The total of 2988 galaxies comprise the full catalogue (Full\_cat) in Table \ref{tab:sampletable}}. In Fig.~\ref{fig:centre_pixcel_bpt} we show the distribution of [O\textsc{iii}]/H$\beta$ versus [N\textsc{ii}]/H$\alpha$ in the central spaxel spectrum of the SAMI sample which have all emission-line S/N > 2. \update{In Fig.~\ref{fig:centre_pixcel_bpt}, we also plot the Kauffmann line which separates SF galaxies from AGN/LINER \citep[the solid blue curve,][]{Kauffmann2003}, the theoretical maximal star-formation line \citep[the red dashed curve,][]{Kewley_2001} and the Seyfert-LINER line which separates Seyferts from LINERs \citep[the green dot-dashed line,][]{Kewley2006}}. Using the central spaxel itself is not sufficient to classify whether the whole galaxy is SF or has LINER/AGN features. To \update{better classify galaxies in Full\_cat, we use the BPT diagram for every spaxel in each galaxy where the emission lines have S/N > 2}. \update{Two examples are shown in} Fig.~\ref{fig:SF_bpt}. The top panel is a SF \update{galaxy} with most spaxels beneath the Kauffmann line while the bottom panel is a galaxy with a SF centre with LINER/shock features in the outskirts. Therefore, we apply a classification to galaxies with central spaxels beneath the Kauffmann line \update{(i.e. SF)}: galaxies \update{that} have more than 2/3 of the spaxels beneath the Kauffmann line are pure SF (p\_SF) galaxies; galaxies \update{that} have more than 2/3 of galaxy spaxels in the composite/LINER region are the central SF (c\_SF) galaxies. With this classification, Fig.~\ref{fig:SF_bpt} top panel \update{shows} a p\_SF galaxy and the bottom panel \update{shows} a c\_SF galaxy. \update{The c\_SF galaxies have SF centres and extended LINER/AGN features. For galaxies with central spectrum above the Kauffmann line: galaxies that have more than 2/3 of the spaxels above the Kewley line are classified as pure AGN; the rest are classified as central LINER/AGN (c\_L/A) galaxies that have LINER/AGN centres with extended star-formation discs.}

Since studying the spatially-resolved star-formation properties of galaxies requires SF galaxies, we use the sample selection criteria of \citet{Schaefer2017} \update{on the the Full\_cat of 2988 galaxies.} SAMI IFU observations have a 15$''$ diameter field of view. To reduce the effect of hexabundles with a finite aperture on measuring the spatial distribution of star-formation, we remove galaxies with effective radii greater than 15$^{\prime\prime}$ (99 galaxies excluded). Edge-on galaxies will hide spatial information and will increase the uncertainty when calculating spatial star-formation properties, so we reject galaxies with ellipticity values greater than 0.7 (another 307 galaxies excluded). We select galaxies with seeing/$R_\mathrm{e} <$ 0.75 and seeing $<$ 4$''$ to reduce the effect of beam smearing on small galaxies (another 714 galaxies excluded). Finally, we exclude 783 galaxies with absorption-corrected H$\alpha$ equivalent widths ($EW_\mathrm{H\alpha}$, integrated over the SAMI cube) less than 1\,\AA, \update{as these are unlikely to have significant star formation}. 121 pure AGN galaxies are excluded as they do not contain any SF spaxels. \update{For the galaxies that have LINER/AGN features and some star formation (c\_SF and c\_L/A galaxies), we correct the H$\alpha$ emission to remove the flux that is not due to star formation (see Section~\ref{sec:star_forming} for details). Once this is done, we remove 186 galaxies that have corrected log(sSFR/yr$^{-1}$) $\le$ $-11.25$.} \update{After all these exclusion, there are 778 galaxies left and form the full sample (Full\_sam) of this work. }The results of these selections are shown in Table~\ref{tab:sampletable}.

\update{In the Full\_sam, 719 galaxies are in the SF sub-sample (Full\_SF). In Full\_SF, 653 galaxies are in the p\_SF sub-sample and 66 galaxies are in the c\_SF sub-sample. The remaining 59 galaxies are in the c$\_$L/A sub-sample which are also listed in Table~\ref{tab:sampletable}.} In the Full\_sam, 649 galaxies are in the SAMI-GAMA catalogue and 129 galaxies are in the SAMI-Cluster catalogue. With 649 GAMA region SF galaxies, the sample size is doubled compared to \citet{schaefer2019}, which used 325 galaxies from SAMI data v0.9.1. A Kolmogorov-Smirnov (K-S) test is applied to compare the Full\_sam with that of \citet{schaefer2019}. The stellar mass ($M_{\star}$) distributions have a $p$-value=0.75 which shows no significant difference.

\subsection{Environmental metrics}
\label{sec:envirmetric}

The SAMI GAMA-region objects cover almost the entire range of environments found in the local Universe (apart from rich clusters) from isolated galaxies to galaxy groups. The GAMA Galaxy Group Catalogue \citep[][]{Robotham2011} is built on a friends-of-friends (FoF) algorithm and halo mass ($M_{200}$) is defined as the mass of a spherical halo with a mean density that is 200 times the critical cosmic density at the halo redshift \update{(the corresponding radius is defined as $R_{200}$)}. The SAMI GAMA-region sample predominantly contains galaxies residing in groups with halo masses $M_{200}$ less than 10$^{14}$ $M_{\odot}$. \update{Centrals and satellites are classified in the GAMA catalogue as well by the FoF algorithm}. The addition of the SAMI cluster sample extends the halo mass range to understand the suppression of star-formation in the highest-density regions. The cluster virial masses should be multiplied by 1.25 in order to match GAMA halo masses \citep[]{Owers2017}. \update{For cluster galaxies, we are mostly considering satellites as the centrals are already quenched. } 

Based on the halo masses, we divide our Full\_sam into four bins: \\1. Ungrouped galaxies (not classified in a group galaxy in the GAMA Galaxy Group Catalogue version 10) \\ 2. Low-mass group galaxies ($M_{200}$ $\leq 10^{12.5} M_{\odot}$) \\ 3. High-mass group galaxies ($M_{200}$ within $10^{12.5-14} M_{\odot}$)\\ 4. Cluster galaxies ($M_{200}$ $> 10^{14} M_{\odot}$)

Many of the low-mass groups have low multiplicity (i.e. pairs) and so have large halo mass uncertainties, where log$_{10}[M_\mathrm{err}/(h^{-1}M_{\odot})]$=1.0$\mathrm{-0.43log_{10}}(N_\mathrm{{FoF}}$) ($N_\mathrm{FoF}$ is the number of member galaxies from the FoF algorithm; \citealt{Robotham2011}). As a result, the ungrouped and low-mass group populations have considerable overlap. In our analysis below we will sometimes combine these two populations as low density environments, when comparing to high-mass groups and clusters.

Along with the global environment each galaxy resides in, we also use the local galaxy density, which is parameterized as the fifth-nearest-neighbour surface density ($\Sigma_{5}$/Mpc$^{2}$; \citealp{Brough_2013,Brough_2017}). The surface density is defined using the projected comoving distance to the fifth nearest neighbour (d$_{5}$) within a velocity range of $\pm$500 km s$^{-1}$ within a pseudo-volume-limited density defining population that have been observed spectroscopically in the GAMA \citep[][]{Brough_2013, Hopkins_2013} regions and clusters \citep[]{Owers2017}: $\Sigma_{5}$ = 5/$\pi d_{5}^{2}$. The density-defining population has absolute SDSS Petrosian magnitude $M_\mathrm{r} < M_\mathrm{r,limit} -Q_\mathrm{z}$ ($M_\mathrm{r,limit}$ = $-20.0$ mag, $Q_\mathrm{z}$ = 1.03; \citealp{Loveday2015}). Similarly to halo masses, we also define four bins in local density: $\Sigma_{5}$ between 0-0.1, 0.1-1, 1-10 and >10 Mpc$^{-2}$.

We also consider projected phase-space diagram which shows both galaxy velocity and radius relative to group/cluster centre. Several studies have connected the star-formation of galaxy populations to their location in phase-space diagrams \citep[e.g.][]{Hernandez2014, Haines2015, Barsanti2018}. The projected velocity over velocity dispersion ($V/\sigma$) and projected distance from the centre of the halo \update{normalized with halo radius $R_{200}$ (r/$R_{200}$)} for clusters are obtained from the SAMI cluster catalogue \citep[][]{Croom_2021, Owers2017}. We use the median redshift as the systemic redshift in groups with redshift of individual galaxies from GAMA catalogue to calculate $V/\sigma$ in groups. The $r/R_{200}$ for groups is calculated from the group velocity dispersion and redshift as described in \citet[]{Finn_2005}.

\section{Star-formation properties and stellar population measurements} \label{sec:star_forming}

We use the dust-corrected H$\alpha$ emission-line fits to calculate SFRs and analyse the radial distribution of star-formation in each galaxy. The H$\alpha$ flux is corrected for LINER/AGN contamination in our sample for central SF galaxies (c\_SF) and central LINER/AGN galaxies (c\_L/A). We expand on these points in detail below.

\subsection{Specific SFRs for SF galaxies}\label{sec:sfr}

The dust-correctted H$\alpha$ is used to calculate sSFR to avoid dust obscuration along the line-of-sight; our correction is based on the dust extinction law of \citet{Cardelli1989}. The attenuation uses the deviation of the Balmer Decrement ($BD$; the ratio ${f_\mathrm{H\alpha}/f_\mathrm{H\beta}}$) from the theoretical value of 2.86 for case B recombination, $n_\mathrm{e} = 10^2\,\mathrm{cm}^2$ and $T = 10^4\,\mathrm{K}$ \citep[][]{Groves_2011}. Because of the weak H$\beta$ luminosity in the outskirts of galaxies, we only apply the dust correction when the S/N of H$\beta$ is larger than 3 per \AA . The SFR is calculated from the total H$\alpha$ luminosity (L$_\mathrm{H\alpha}$) using the \citet{kennicutt1998global} relation with a \citet{chabrier2003galactic} Initial Mass Function (IMF), as follows:

\begin{gather}
    \mathrm{SFR}=\frac{L_\mathrm{H\alpha}(W)}{2.16\times10^{34}}{M_\odot\,yr^{-1}}.
	\label{eq:SFR}
\end{gather}

For pure SF (p\_SF) galaxies, we use the total H$\alpha$ flux to calculate SFR as we assume that all the H$\alpha$ flux is directly associated with star-formation. We will discuss the H$\alpha$ flux calculation for composite/LINER/AGN spaxels in the following Section~\ref{sec:agncorretion}. The sSFR is then calculated as SFR/$M_{\star}$.

\begin{figure*}
	\includegraphics[scale=0.65]{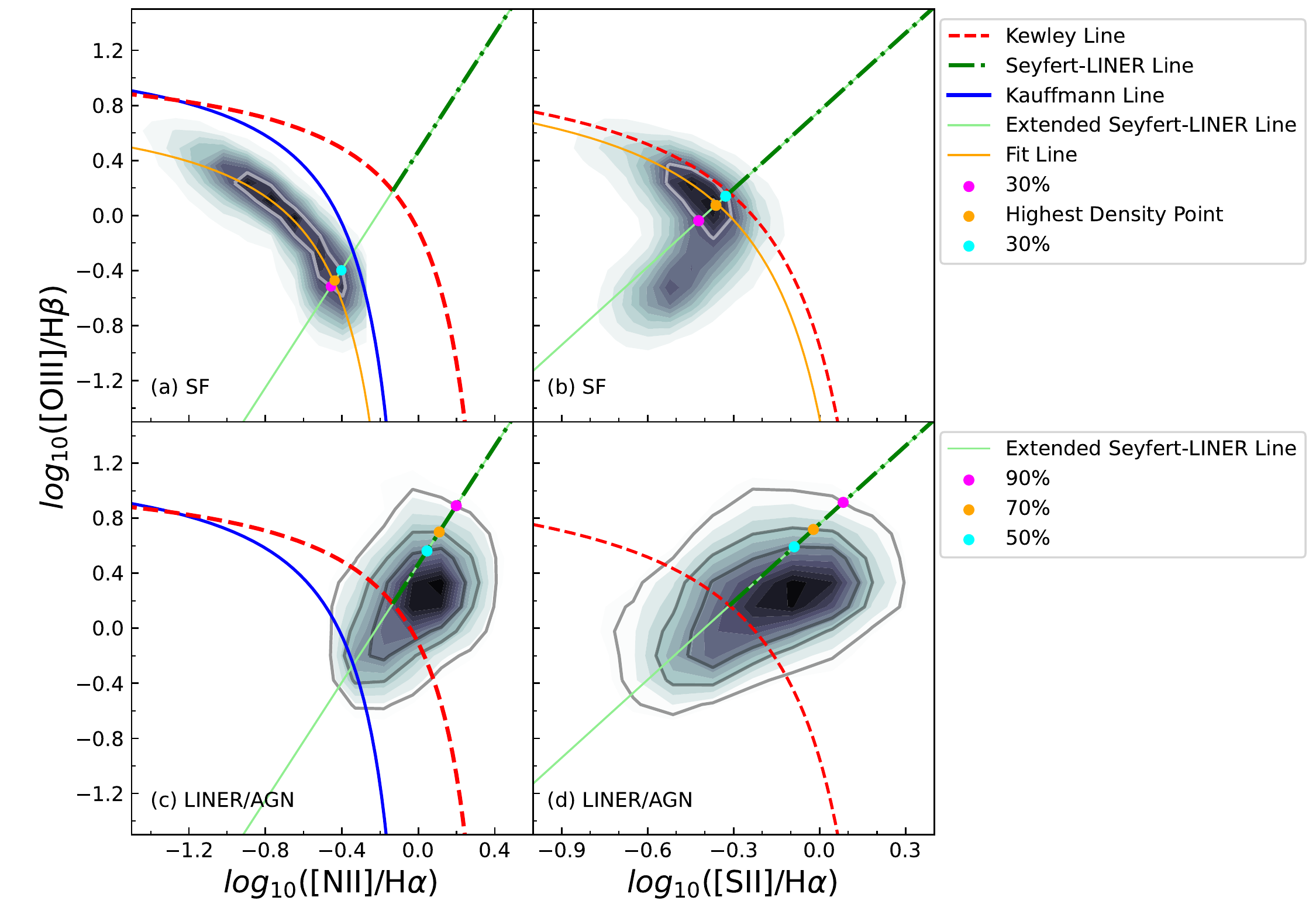}
    \caption{We show our 100 per cent SF and 100 per cent LINER/AGN points on 2 types of BPT diagrams using the central spaxels of each galaxy in the full SAMI sample. (a) and (c) show the BPT contour diagram of [O\textsc{iii}]/H$\beta$ vs [N\textsc{ii}]/H$\alpha$. (b) and (d) show the BPT contour diagram of [O\textsc{iii}]/H$\beta$ vs [S\textsc{ii}]/H$\alpha$. The light green line is the extended LINER line. In the top two panels, we plot all central spaxel emission-line ratios beneath the blue (\citealt{Kauffmann2003}) line. The highest density (orange point) indicates 100 per cent star-formation emission (SF point). The 30 per cent density cross-points (magenta and cyan) are used for the uncertainty range. In the bottom two panels, we show all central spaxel emission-line ratios above the blue (\citealt{Kauffmann2003}) line as the LINER/AGN/composite galaxies in the full SAMI sample. The contour lines are 50 per cent, 70 per cent and 90 per cent of the sample. The 70 per cent contour line cross-point (orange point) indicates the 100 per cent LINER/AGN-like emission (L/A point). The 50 per cent (cyan) and 90 per cent (magenta) cross-points are for the uncertainty ranges.}
    \label{fig:bpt_con}
\end{figure*}

\begin{figure}
	\includegraphics[width=\columnwidth]{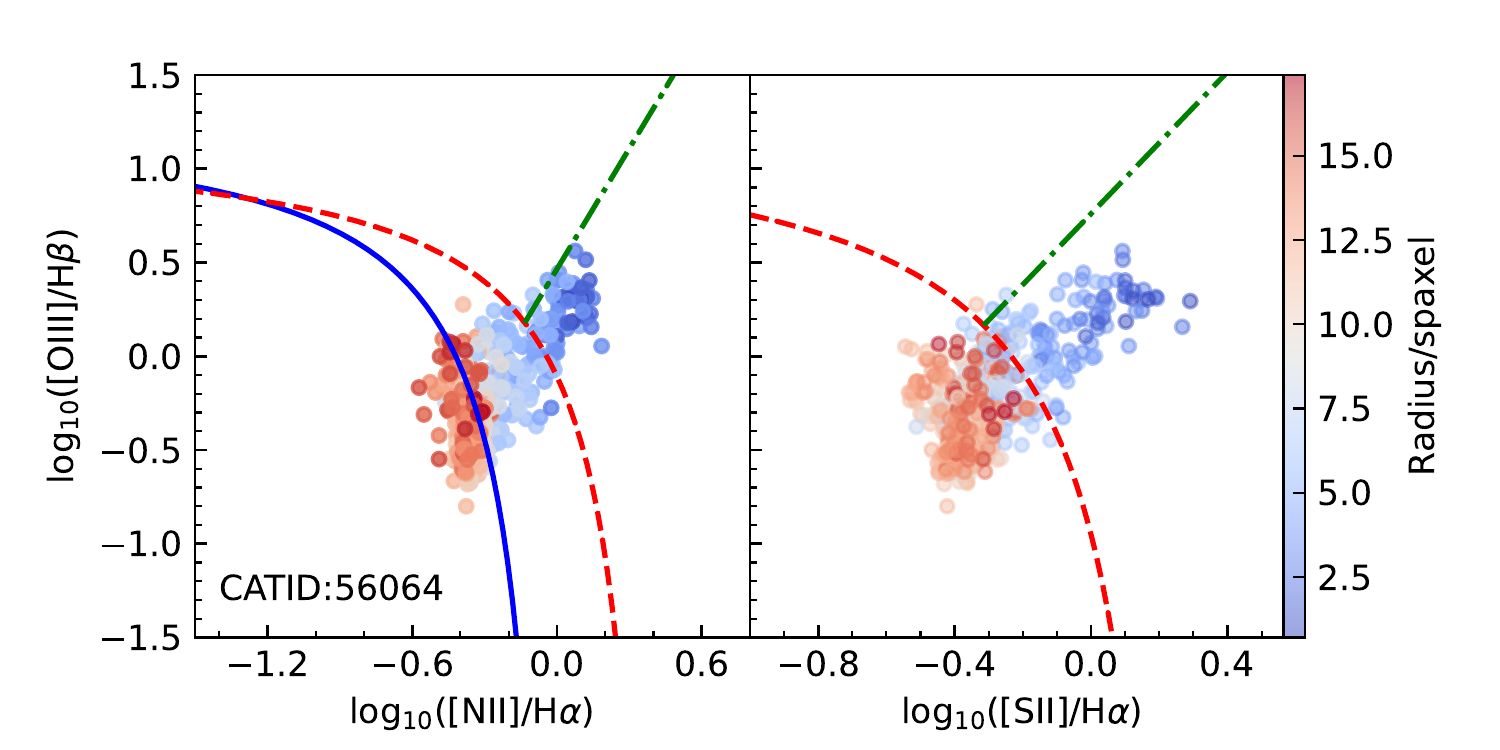}
    \caption{Spatially-resolved BPT diagrams of a galaxy with central LINER/AGN components and SF disc (c$\_$L/A). The left panel shows the [O\textsc{iii}]/H$\beta$ versus [N\textsc{ii}]/H$\alpha$ and the right shows [O\textsc{iii}]/H$\beta$ versus [S\textsc{ii}]/H$\alpha$ for the spaxels of this galaxy, which all have emission-line S/N larger than 2, colour-coded by radius in the galaxy. The blue, red, and green lines are the same as in Fig.~\ref{fig:centre_pixcel_bpt}.}
    \label{fig:agn_ratio_example_2}
\end{figure}

\begin{figure}
	\includegraphics[width=\columnwidth]{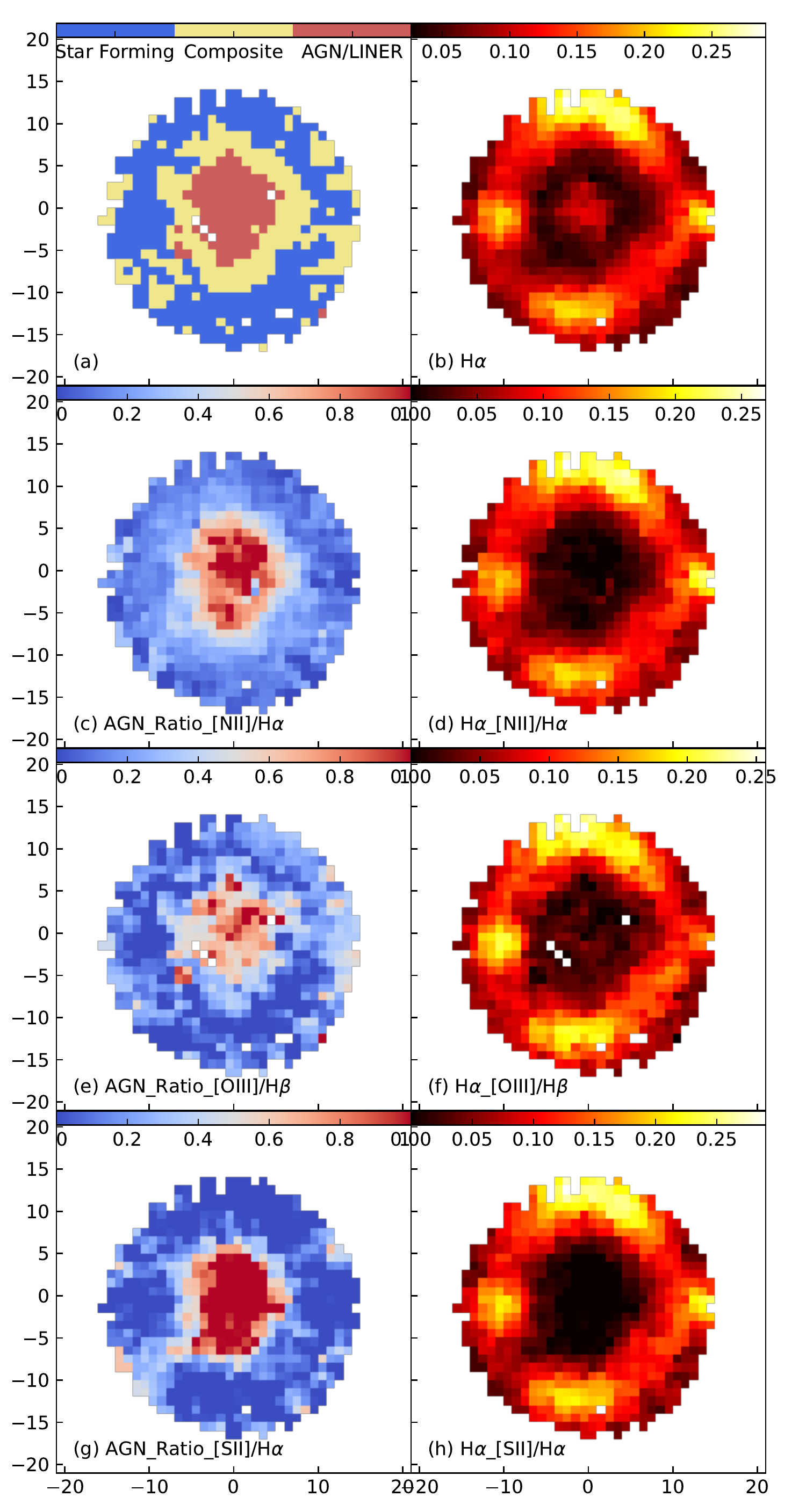}
    \caption{Maps of a c$\_$L/A galaxy \update{demonstrating classification and AGN/LINER correction.} Panel (a) shows the distribution of the galaxy spaxels located in the BPT diagram with blue for star-formation, green for composite and red for LINER/AGN. Panel (b) shows the original H$\alpha$ map. Panels (c, e, g) are LINER/AGN ratios calculated using the BPT diagram in [N\textsc{ii}]/H$\alpha$, [O\textsc{iii}]/H$\beta$ and [S\textsc{ii}]/H$\alpha$ with blue colour for 100 per cent LINER/AGN emission and red colour for 100 per cent star-formation emission. Panels (d, f, h) show the resulting LINER/AGN corrected H$\alpha$ map.}
    \label{fig:agn_ratio_example}
\end{figure}

\subsection{H$\alpha$ emission-line correction for LINER/AGN-like galaxies}\label{sec:agncorretion}

\citet{schaefer2019} argued that excluding LINERs/AGNs would be unlikely to bias the relationship with the environment but could introduce a bias as a function of $M_{\star}$. The majority of LINER/AGN-like galaxies have stellar masses in the $10^{10-11} M_{\odot}$ range. However, to fully explore the impact of AGN, we include them in our analysis and make a correction for the H$\alpha$ flux due to LINER/AGN emission \citep[][]{Davies2014}. We generally use LINER/AGN in this paper to mean any non SF emission (e.g. including winds or shocks). This correction has the advantage of allowing us to include galaxies with SF discs, but passive centres that may have some weak central LINER/AGN-like emission. \citet{Belfiore2016} used the [S\textsc{ii}]/H$\alpha$ ratio to quantify the residual star-formation in galaxies with quiescent central like regions with typical line ratios of \((\frac{[S\textsc{ii}]}{\mathrm{H\alpha}})_\mathrm{SF}\)=0.4 and \((\frac{[S\textsc{ii}]}{\mathrm{H\alpha}})_\mathrm{LIER}\)=1.0 (LIER for low-ionization emission-line region) using the BPT diagram. We use an approach qualitatively similar to \citet[]{Belfiore2016}, but consider other line ratios as well. 

We take the [N\textsc{ii}]/H$\alpha$, [O\textsc{iii}]/H$\beta$ and [S\textsc{ii}]/H$\alpha$ emission-line ratios to calculate how much H$\alpha$ is contributed by star-formation or LINER/AGN emission for each galaxy spaxel. If a galaxy has a LINER feature, the emission-line ratio distribution is more horizontal, then the galaxy is sensitive to the [N\textsc{ii}]/H$\alpha$ and [S\textsc{ii}]/H$\alpha$. For a Seyfert galaxy, the emission-line ratio distribution is more vertical, the galaxy is sensitive to [O\textsc{iii}]/H$\beta$. To test the three emission-line ratios, we need to define regions in the BPT diagram corresponding to 100 per cent star-formation (SF point) and 100 per cent LINER/AGN emission (L/A point). \update{Based on whether galaxies have more LINER features or AGN features, the 100 per cent LINER/AGN emission can vary. Therefore, to find the most representative ratio for all galaxies in Fig.~\ref{fig:centre_pixcel_bpt} LINER/AGN region, rather than defining a new line, we choose to follow the Seyfert-LINER line (green) as it provides an empirical division between Seyferts and LINERs \citep[][]{Kewley2006}. If we extended the Seyfert-LINER line, it intersects with the SF locus and does a good job of following the overall locus of points as they move from SF to composite to LINER/AGN regions.} Therefore, the following assumptions are made: 1) the SF point is on the SF galaxy locus; 2) the L/A point is on the composite/LINER/AGN galaxy locus; 3) both the SF point and L/A point are on the \update{extended} Seyfert-LINER line. With these assumptions, the SAMI sample is classified into SF galaxies (beneath the Kauffmann line) and composite/LINER/AGN galaxies (above the Kauffmann line) using the location of their central spaxel spectra in the BPT diagram. These are plotted separately in Fig.~\ref{fig:bpt_con}.  

To find the 100 per cent SF point, the same mathematical form as the \update{Kewley} line is used and moved to the highest density SF galaxy locus \update{(orange line in Fig.~\ref{fig:bpt_con} a: \(\log_{10}(\mathrm{[O\textsc{iii}]}/{\mathrm{H\beta}})\)=0.61/\(\log_{10}(\mathrm{[N\textsc{ii}]}/{\mathrm{H\alpha}})\)+0.9006; orange line in Fig.~\ref{fig:bpt_con} b: \(\log_{10}(\mathrm{[O\textsc{iii}]}/{\mathrm{H\beta}})\)=0.72/[\(\log_{10}(\mathrm{[S\textsc{ii}]}/{\mathrm{H\alpha}})\)-0.265]+1.24)}. As choosing 100 per cent SF points (orange points, cross points of the extended Seyfert-LINER line) are manually selected, the 30 per cent density contour line is chosen as an uncertainty limit (the crossing points in magenta and cyan in panel a and b). \update{For LINER/AGN galaxies, there is not a well-defined locus that is 100 per cent AGN (unlike the SF locus) and the Seyfert-LINER line does not pass most density region. Compromising,} the crossing points of the 50 per cent, 70 per cent and 90 per cent of the contour lines are plotted to find the 100 per cent L/A points. Although the 70 per cent contour line crossing points (orange) are not at the highest density point for [O\textsc{iii}]/H$\beta$, they represent [N\textsc{ii}]/H$\alpha$ and [S\textsc{ii}]/H$\alpha$ well and are chosen as the 100 per cent L/A points. The 50 per cent and 90 per cent contour line crossing points (cyan and magenta points in panels c and d) are chosen to define the uncertainty. We check that changing these line ratios in our uncertainty ranges does not significantly affect the conclusions of the paper, as discussed further in Section~\ref{sec:cindex_for_lA}.

The SF points and L/A points give us 3 emission-line ratio ([N\textsc{ii}]/H$\alpha$, [O\textsc{iii}]/H$\beta$ and [S\textsc{ii}]/H$\alpha$) boundaries. With emission-line ratio boundaries, we can define a linear scale from 0 to 1. This scale is called LINER/AGN ratio ($R_\mathrm{L/A}$). $R_\mathrm{L/A}$ = 1 represents pure LINER/AGN while $R_\mathrm{L/A}$ = 0 represents pure SF. To summarise, the orange points represent the 100 per cent SF and 100 per cent L/A points; the cyan/magenta points indicate the uncertainty. In this way, for each spaxel, the H$\alpha$ flux contributed by star-formation can be calculated as:

\begin{gather}
    F_\mathrm{H\alpha,SF}=f_\mathrm{H\alpha}\times(1-R_\mathrm{L/A}),
	\label{eq:ratio_agn}
\end{gather}
\begin{gather}
    R_\mathrm{L/A}=\frac{1-\log_{10}([N\textsc{ii}]/H\alpha)}{\log_{10}([N\textsc{ii}]/H\alpha)_\mathrm{L/A}-\log_{10}([N\textsc{ii}]/H\alpha)_\mathrm{SF}}.
	\label{eq:ratio_agn2}
\end{gather}

In Equation.~(\ref{eq:ratio_agn2}), [N\textsc{ii}]/H$\alpha$ can be replaced by [O\textsc{iii}]/H$\beta$ and [S\textsc{ii}]/H$\alpha$. The emission-line ratios of full star-formation and full LINER/AGN (L/A) are \(\log_{10}(\frac{\mathrm{[N\textsc{ii}]}}{\mathrm{H\alpha}})_\mathrm{SF}\) = $-0.438$, \(\log_{10}(\frac{\mathrm{[N\textsc{ii}]}}{\mathrm{H\alpha}})_\mathrm{L/A}\) = 0.103; \(\log_{10}(\frac{\mathrm{[S\textsc{ii}]}}{\mathrm{H\alpha}})_\mathrm{SF}\) = $-0.362$, \(\log_{10}(\frac{\mathrm{[S\textsc{ii}]}}{\mathrm{H\alpha}})_\mathrm{L/A}\) = $-0.047$; \(\log_{10}(\frac{\mathrm{[O\textsc{iii}]}}{\mathrm{H\beta}})_\mathrm{SF}\) = $-0.473$,  \(\log_{10}(\frac{\mathrm{[O\textsc{iii}]}}{\mathrm{H\beta}})_\mathrm{L/A}\) = 0.686. As we have calculated how much H$\alpha$ flux is contributed by star-formation, we can use the corrected H$\alpha$ flux to calculate SFR. 

A galaxy with central LINER/AGN-like emission is shown in Fig.~\ref{fig:agn_ratio_example_2} and Fig.~\ref{fig:agn_ratio_example}. Fig.~\ref{fig:agn_ratio_example_2} shows the spatially-resolved BPT diagrams for the example galaxy. The classification map (Fig.~\ref{fig:agn_ratio_example}a) shows that this galaxy has a SF disc at large radius. The residual H$\alpha$ flux from star-formation is calculated using $R_\mathrm{L/A}$. Fig.~\ref{fig:agn_ratio_example}(b) shows the original H$\alpha$ map. Fig.~\ref{fig:agn_ratio_example}(c, e, g) show the three $R_\mathrm{L/A}$ ratios, red for 100 per cent AGN and blue for 100 per cent star-formation. This galaxy has a LINER feature, so the emission-line is less sensitive to the [O\textsc{iii}]/H$\beta$ compared to [N\textsc{ii}]/H$\alpha$ and [S\textsc{ii}]/H$\alpha$. The $R_\mathrm{L/A}$ calculated by 
[N\textsc{ii}]/H$\alpha$ and [S\textsc{ii}]/H$\alpha$ is high in the centre. Fig.~\ref{fig:agn_ratio_example}(d, f, h) show the residual star-formation contributed H$\alpha$ maps. The sSFR is then be calculated with residual H$\alpha$ flux for c\_L/A and c\_SF galaxies. 

\begin{figure}
	\includegraphics[width=7cm]{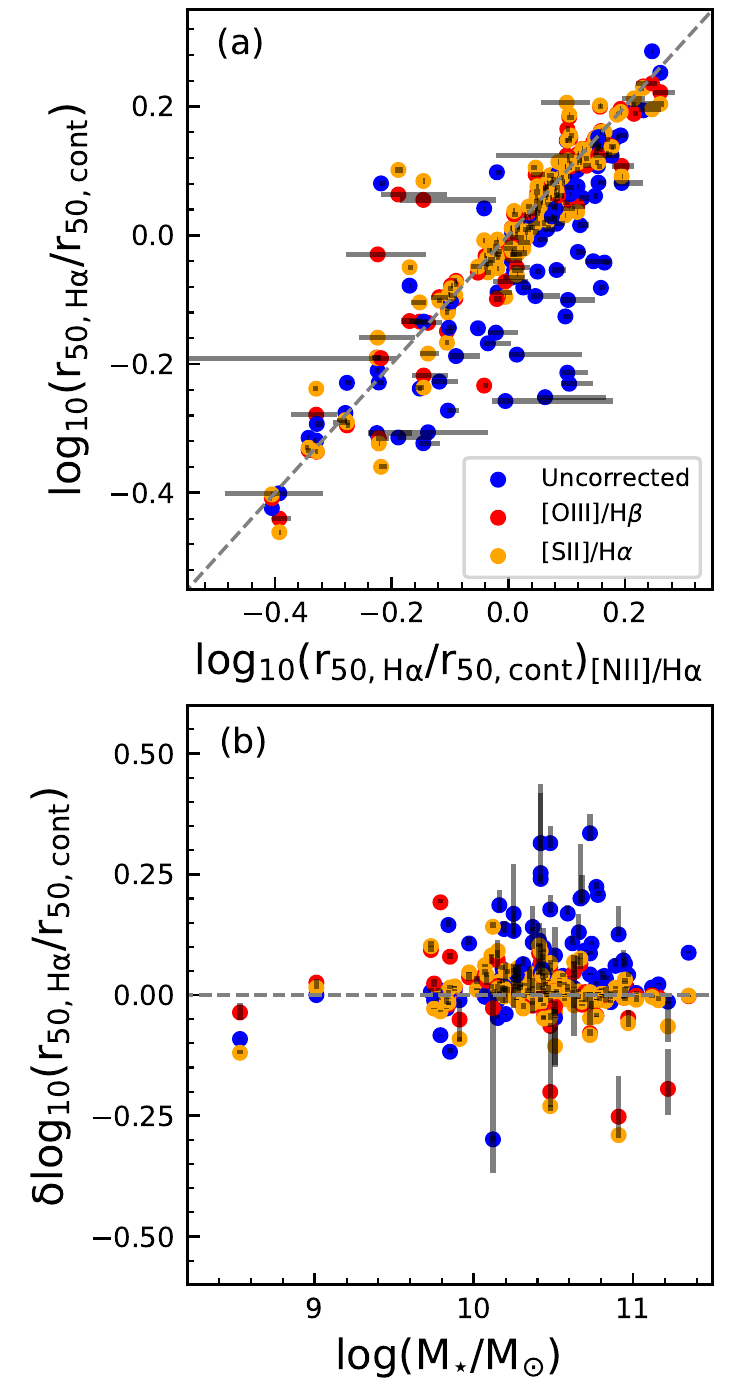}
    \caption{We show differences of the $C$-index before and after 3 types of AGN/LINER correction. The top panel shows uncorrected $C$-index(blue), corrected by [O\textsc{iii}]/H$\beta$ ratio (red) and corrected by [S\textsc{ii}]/H$\alpha$ ratio (orange) versus $C$-index corrected by [N\textsc{ii}]/H$\alpha$ ratio with their uncertainties. Note, for the uncorrected $C$-index, the error-bar is for [N\textsc{ii}]/H$\alpha$. \update{The bottom panel shows the difference between the uncorrected $C$-index (blue); $C$-index use H$\alpha$ corrected by [O\textsc{iii}]/H$\beta$ (red); corrected by [S\textsc{ii}]/H$\alpha$ (orange) and corrected by [N\textsc{ii}]/H$\alpha$ versus $M_{\star}$. As for the top panel, the error-bars on the uncorrected $C$-index is for $C$-index corrected by [N\textsc{ii}]/H$\alpha$.} It shows that taking into account the LINER/AGN-like galaxies can increase the $C$-index by up to a factor 0.25. The [N\textsc{ii}]/H$\alpha$ corrected $C$-index is used in our analysis.}
    \label{fig:agncindex_compare}
\end{figure}

\subsection{The spatial distribution of star-formation}
\label{sec:cindex_for_lA}
Having removed LINER/AGN contribution to the H$\alpha$ flux, we can use our corrected H$\alpha$ maps to study the spatial distribution of star-formation. Following \citet{Schaefer2017}, we introduce the ratio of $r_\mathrm{50,H\alpha}/r_\mathrm{50,cont}$ which compares the half-light radius of H$\alpha$ ($r_\mathrm{50,H\alpha}$) and half-light radius of the $r$-band continuum ($r_\mathrm{50,cont}$). The star-formation concentration index ($C$-index) is defined as $\log_{10}(r_\mathrm{50,H\alpha}/r_\mathrm{50,cont})$ and it indicates the ongoing \update{star-formation} distribution in galaxies.

To understand the systematic uncertainty of calculating the $C$-index ratio for c\_SF galaxies and c\_L/A galaxies, the three emission-line ratio corrected $C$-index measurements are compared with the uncorrected measurements in Fig.~\ref{fig:agncindex_compare}. Fig.~\ref{fig:agncindex_compare}a shows $C$-index uncorrected (blue), corrected by [O\textsc{iii}]/H$\beta$ (red) and corrected by [S\textsc{ii}]/H$\alpha$ (orange) versus $C$-index corrected by [N\textsc{ii}]/H$\alpha$. The error-bars are calculated from the maximum/minimum range allowed by the SF point and L/A point. For the uncorrected (blue), the grey bar shows the uncertainties calculated by [N\textsc{ii}]/H$\alpha$. 

Most of the galaxies follow the one-to-one reference line but with some outliers. Fig.~\ref{fig:agncindex_compare} shows most galaxies have small error-bars (grey, some error-bars are too small to see). Larger error-bars mean LINER/AGN corrections for those galaxies may overestimate/underestimate the H$\alpha$ flux LINER/AGN contribution. Fig.~\ref{fig:agncindex_compare}b shows the difference between \update{the [N\textsc{ii}]/H$\alpha$ correction and no correction (blue points); the [O\textsc{iii}]/H$\beta$ correction (red points); the [S\textsc{ii}]/H$\alpha$ correction (orange points) versus $M_{\star}$. The uncorrected $C$-index} are slightly offset as expected with uncorrected H$\alpha$ flux. \update{The correction we make tends to make star-formation less concentrated}. The $C$-index will be larger than before applying the LINER/AGN correction. This shows that taking into account the LINER/AGN-like galaxies can increase the $C$-index by up to 0.25. [N\textsc{ii}]/H$\alpha$ has on average the highest S/N, so we adopt it as the default correction. We calculate the Spearman rank correlations between all four $C$-indices, and the [N\textsc{ii}]/H$\alpha$ correction has the best correlation of 0.924 with [S\textsc{ii}]/H$\alpha$, 0.947 with [O\textsc{iii}]/H$\beta$ and 0.763 with uncorrected. The high Spearman correlation with the $C$-indices derived from [S\textsc{ii}]/H$\alpha$ and [O\textsc{iii}]/H$\beta$ confirms that we are not biasing our results by using [N\textsc{ii}]/H$\alpha$. As a result, the [N\textsc{ii}]/H$\alpha$ corrected $C$-index is used in our analysis. To further investigate quenching time-scales, we will introduce stellar population measurements in the next section.

\subsection{Stellar population age measurements and $D_\mathrm{n}$4000, $H\delta_\mathrm{A}$ indices }
\label{sec:dn4000}
We use the full-spectral fitting code pPXF which infers the best-fit stellar population parameters in each spectrum to calculate stellar ages. To increase the S/N of our age measurements, we use SAMI sector binning data. By using the $\mathrm{FIND\_GALAXY}$ routine \citep[][]{cappellari2002}, 5 ellipses are defined by collapsing the SAMI cubes in the wavelength direction. With those ellipses, SAMI cubes are binned into five linearly spaced elliptical annuli \citep[e.g.][]{Scott2018,Croom_2021}. Then, the SAMI sector bins azimuthally subdivide each of the annular bins into eight equal area regions. The sector bins have S/N $\geq$ 10.

The [$\alpha$/Fe] enhanced MILES model library \citep{Vazdkes2015} is used with templates that span a range of ages between 0.03 and 14 Gyr, metallicities ([M/H]) between $-2.3$ and +0.4 dex and $\alpha$-enhancement ([$\alpha$/Fe]) between 0.0 and 0.4 dex. The total number of SSP templates used during the fitting is 288. The fit covers the full SAMI wavelength range. To avoid masking emission-line regions, we include templates for the ionised emission-lines corresponding to the chemical species H$\alpha$, H$\beta$, [N\textsc{ii}], [S\textsc{ii}], [O\textsc{i}] and [O\textsc{iii}]. The kinematics of the stellar and emission-line templates are fitted simultaneously but are not forced to take the same values. We recover three separate kinematic solutions; for the stellar component, the emission-line templates corresponding to the Balmer series (H$\alpha$ and H$\beta$) and the templates corresponding to the remaining emission-lines. During the fit, we correct for different continuum shapes between model and data by fitting for a multiplicative polynomial of order 10. 

With the full spectral fits, we calculate the mass-weighted ($Age_\mathrm{M}$) and
light-weighted stellar ages ($Age_\mathrm{L}$) for our galaxies. Following \citet[]{Mcdermid2015}, the $Age_\mathrm{M}$ of each spectrum can be calculated \update{from SFH} by   
\begin{gather}
    \log({Age}_\mathrm{SFH})=\frac{\Sigma w_\mathrm{i} \log(t_\mathrm{SSP,i})}{\Sigma w_\mathrm{i}}.
	\label{eq:age equa}
\end{gather}
Where $w_{i}$ are the best-fitting template weights derived from pPXF $i$th
template, which has age $t_\mathrm{SSP,i}$ and metallicity $[Z/H]_\mathrm{SSP,i}$.

Along with galaxy ages, we also derive the $D_\mathrm{n}$4000 and $H\delta_\mathrm{A}$ indices which are direct measurements from the SAMI cubes, to support our age measurements. The $D_\mathrm{n}$4000 index is well known to be a good stellar population age indicator for intermediate to old stellar populations \citep[e.g.][]{Kauffmann2003}. The feature is caused by a large number of spectral lines occurring around 4000 \AA, mostly due to ionized metals. The $D4000$ index becomes stronger in old metal rich stellar populations. The narrow band $D_\mathrm{n}$4000 index (3850$-$3950 and 4000$-$4100 \AA) defined by \citet{Balogh_1999} is used in this work. The H$\delta$ absorption line equivalent width is also an indicator of galaxy stellar population age. The $H\delta_\mathrm{A}$ index \citep[][]{Worthey_1994} is sensitive to younger populations compared to $D_\mathrm{n}$4000 since the peak occurs in fast terminated A stars. SAMI blue band spectra fits are pre-fitted by pPXF to remove Balmer emission in the $H\delta_\mathrm{A}$ and $D_\mathrm{n}$4000 wavelength ranges. After fitting the spectra, sector bins that have $H\delta_\mathrm{A}$ error greater than 1 \AA $ $ and $D_\mathrm{n}$4000 index error larger than 0.1 are excluded. We compare the derived age measurements for galaxies in different halo mass intervals in Section~\ref{sec:age radial profile}.
 
\begin{figure*}
	\includegraphics[width=18cm]{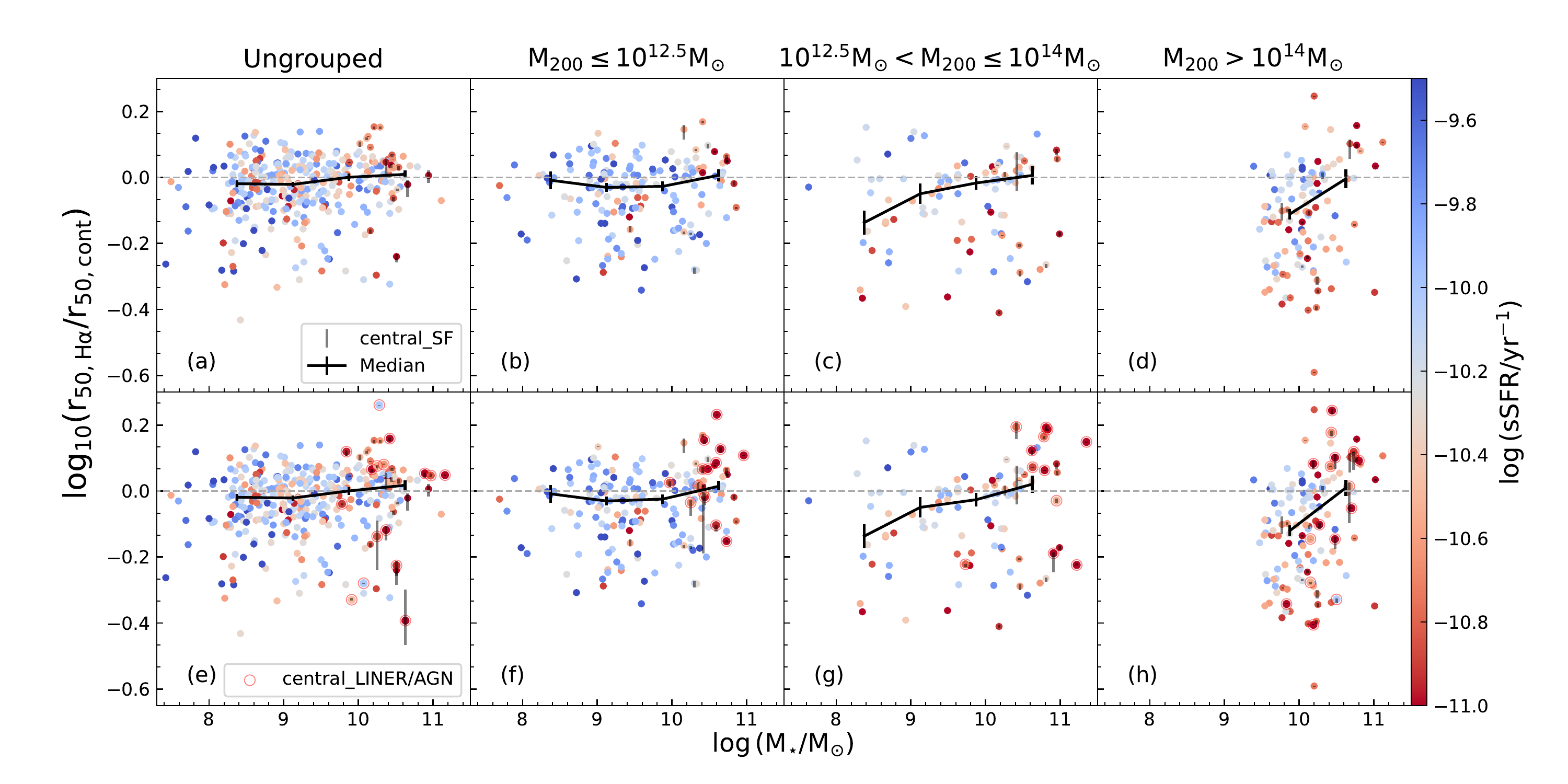}
    \caption{$C$-index as a function of $M_{\star}$ in four different halo mass intervals. From left to right: these are ungrouped galaxies, low-mass group galaxies, high-mass group galaxies and cluster galaxies, colour-coded by sSFR. The top row shows galaxies in the full SF (Full\_SF) sample. The galaxies with error-bars are \update{the c\_SF galaxies}. The error-bars are based on \update{the uncertainty in the LINER/AGN correction}. The bottom row shows galaxies in the Full\_sam including c\_L/A galaxies (red circles) with the error-bars showing the uncertainty due to the LINER/AGN correction. Median $C$-indices in different $M_{\star}$ bins are shown by the black line with standard error of the median as uncertainties. }
    \label{fig:CvsM}
\end{figure*}

\begin{figure}
	\includegraphics[width=7cm]{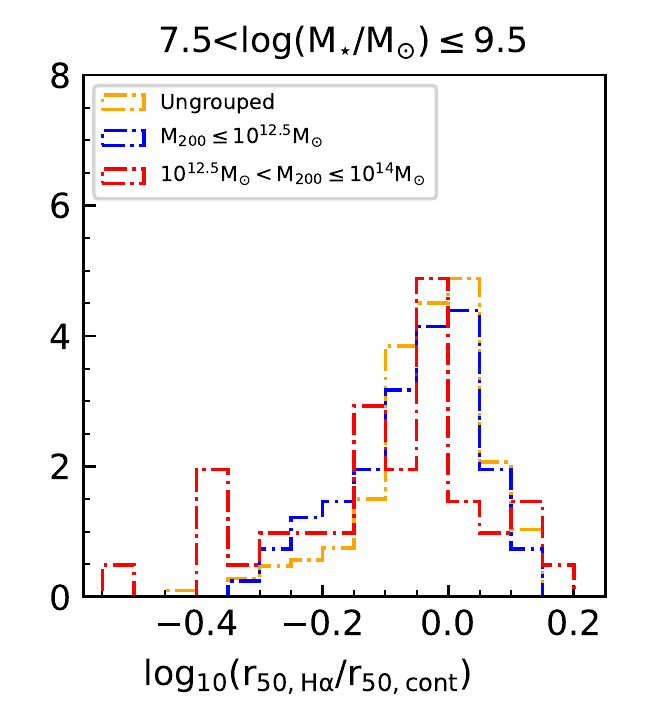}
    \caption{Normalized histogram of $C$-index for galaxies in the Full\_sam in 3 halo mass intervals with stellar mass of 10$^{7.5-9.5}$ $M_{\odot}$. Ungrouped galaxies are in yellow, low-mass groups in blue and high-mass groups in red. For the low $M_{\star}$ bin, there is a shift of $C$-index locus to lower values in high-mass groups compared to ungrouped galaxies and low-mass groups.}
    \label{fig:massbin2}
\end{figure}

\begin{figure}
	\includegraphics[width=7cm]{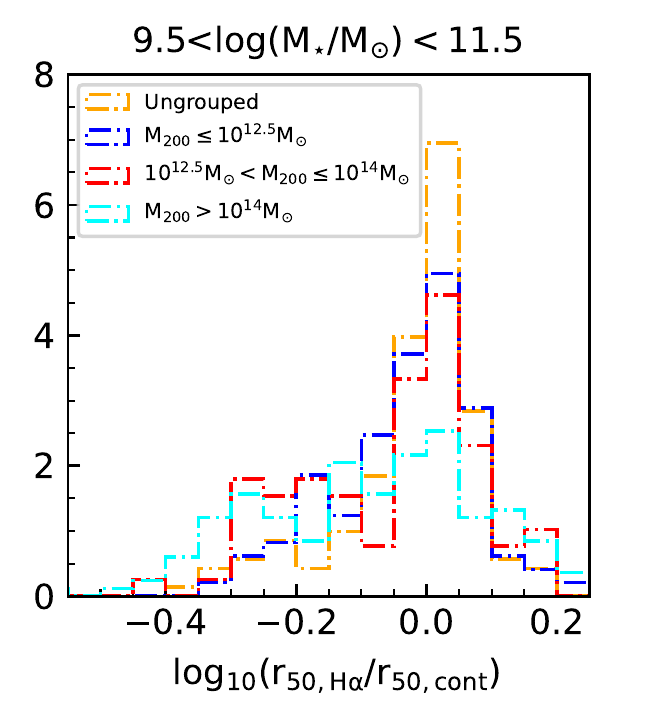}
    \caption{Normalized histogram of $C$-index in 4 halo mass intervals for galaxies with stellar mass of 10$^{9.5-11.5}$ $M_{\odot}$ from the Full\_sam. Ungrouped galaxies are in yellow, low-mass groups in blue, high-mass groups in red and cluster galaxies in cyan. We find the tail of galaxies with concentrated \update{star-formation} increases with higher halo mass.}
    \label{fig:massbin_agn}
\end{figure}

\begin{table}
	\centering
	\caption{The number of galaxies in \update{Full\_sam, Full\_SF, and} c\_L/A sub-sample in each halo mass bin. Number of galaxies in Full\_sam with $M_{\star}$ bin of 10$^{7.5-9.5}$ $M_{\odot}$ and 10$^{9.5-11.5}$ $M_{\odot}$.}
	\label{tab:number_halo}
	\begin{tabular}{lcccc} 
		\hline
		$ $ & ungrouped & Group(low) & Group(high) & Cluster  \\
		$ $ & $ $ & ${\le} 10^{12.5}M_{\odot}$ & $ 10^{12.5-14}M_{\odot}$ &${>} 10^{14}M_{\odot}$\\
		\hline
		Full\_sam & 351 & 179 & 109 & 129 \\
		Full\_SF & 335 & 165 & 107 & 112 \\
		c\_L/A  & 16 &   14 &  12 &  17 \\
		\hline
		10$^{7.5-9.5}M_{\odot}$ & 213 & 82 & 41 & $ $\\
		10$^{9.5-11.5}M_{\odot}$ & 138 & 97 & 78 & 129 \\ 
		\hline
	\end{tabular}
\end{table}

\section{Results: The spatial distribution of star-formation}\label{sec:result_spatialSF}
Our main focus is to understand the suppression of star-formation in high-density regions. We focus on $C$-index to examine the extent of star-formation as a function of stellar mass, shown in different environments. The environments are defined primarily by the host halo mass. Additionally, the fifth-nearest-neighbour density, which shows the local environmental density in which each galaxy resides, is also explored as an additional environmental metric.  

\begin{table}
	\centering
	\caption{Results of the K-S test of $C$-index in the Full\_sam within four halo mass intervals for galaxies with $M_{\star}$ of 10$^{9.5-11.5}$ $M_{\odot}$. \update{The statistic value ($D$) and probability ($p$) values are shown in the table.} We mark significant $p$-values in bold.}
	\label{tab:kstest}
	\begin{tabular}{cccc} 
		\hline
		K$-$S test & Group(low) & Group(high) & Cluster  \\
		$ $ & ${\le} 10^{12.5} M_{\odot}$&$ 10^{12.5-14} M_{\odot}$&${>} 10^{14} M_{\odot}$\\
		\hline
		ungrouped & $D$=0.115, & $D$=0.201, & $D$=0.278, \\
		$ $ &  $p$=0.397 & $p$=\textbf{0.030} & $p$=\textbf{1.025e-05}\\
		\hline
		Group(low) & $ $ & $D$=0.158, & $D$=0.228,\\
		(${\le} 10^{12.5} M_{\odot}$)&$ $& $p$=0.199 & $p$=\textbf{0.003}\\
		\hline
		Group(high) & $ $ & $ $ & $D$=0.149\\
		($ 10^{12.5-14} M_{\odot}$)&$ $& $ $&$p$=0.168\\
		\hline
	\end{tabular}
\end{table}

\subsection{The spatial star-formation distribution in galaxy groups and clusters}

The ratio of $C$-index as a function of $M_{\star}$ is shown in different environments, colour-coded by sSFR in Fig.~\ref{fig:CvsM}. We divide our galaxies into 4 halo mass bins: ungrouped galaxies, low-mass group galaxies, high-mass group galaxies and cluster galaxies (Section~\ref{sec:envirmetric}). The top row shows galaxies in the Full\_SF sample. The galaxies with error-bars are from the central SF (c\_SF) sub-sample, that have been corrected for some non-SF spaxels at large radii. The error-bars are estimated from the correction for non-SF emission (typically shocks for galaxies with central star-formation). There are 335, 165, 107 and 112 SF galaxies in panels a-d respectively. The bottom row of Fig.~\ref{fig:CvsM} includes the \update{Full\_SF} sample and the c\_L/A galaxies (red circles) with error-bars from the non-SF emission correction. An extra 16, 14, 12 and 17 c\_L/A galaxies are added to the previous \update{Full\_SF} sample in panel e-h respectively. These numbers are shown in Table~\ref{tab:number_halo}. Galaxies with $C$-index less than 0 have H$\alpha$ emission that is more compact than their r-band continuum. Galaxies with $C$-index larger than 0 have extended star-formation. The medians in the mass bin are shown by black lines. The uncertainties of the medians are calculated by the standard error of the mean. 

As shown in Fig.~\ref{fig:CvsM}, ungrouped galaxies and galaxies in low halo mass groups tend to have $C$-index close to 0. More massive environments tend to have a larger range of $C$-index values. The c\_L/A galaxies are located over a large range in $C$-index but at higher $M_{\star}$, mostly $>$ $10^{10} M_{\odot}$. By definition, the c\_L/A galaxies are galaxies that have a SF disc with central LINER/AGN emission (an example can be seen in Fig.~\ref{fig:agn_ratio_example}), as a result, 59 per cent of these galaxies have $C$-index $>$ 0. Galaxies with high $C$-index ($>$ 0.1) either show a SF disc with central strong LINER/AGNs or have a broadened star-formation distribution. The uncertainties on $C$-index due to the AGN correction are small in most cases, significantly smaller than the width of the overall distribution in $C$-index. For the low mass end (10$^{9.5-11.5}M_{\odot}$ $M_{\star}$), we can see a lower $C$-index shows out in high-mass groups than ungrouped and low-mass group galaxies. 

Because of the sample selection of SAMI, only galaxies with $M_{\star}$ of 10$^{9.5-11.5}M_{\odot}$ are observed in clusters \citep[][]{Bryant_2015}. To match $M_{\star}$ for different halo mass intervals, our sample is separated into two $M_{\star}$ bins of 10$^{9.5-11.5}M_{\odot}$ and 10$^{7.5-9.5}M_{\odot}$. In the 10$^{9.5-11.5}M_{\odot}$ mass bin, there are 138, 97, 78 and 129 galaxies in the ungrouped region, low-mass groups, high-mass groups and clusters, respectively. In the 10$^{7.5-9.5}M_{\odot}$ mass bin, there are 213, 82 and 41 galaxies in the ungrouped region, low-mass groups and high-mass groups, respectively. The summary of these numbers are shown in Table~\ref{tab:number_halo}.

The histogram of $C$-index for our Full\_sam in the two $M_{\star}$ bins are shown in Fig.~\ref{fig:massbin2} (10$^{7.5-9.5}M_{\odot}$ $M_{\star}$ bin) and Fig.~\ref{fig:massbin_agn} (10$^{9.5-11.5}M_{\odot}$ $M_{\star}$ bin). For the lower-$M_{\star}$ galaxies in Fig.~\ref{fig:massbin2}, the $C$-index locus in high-mass groups is slightly lower than ungrouped and low-mass groups. Fig.~\ref{fig:massbin_agn} shows the $C$-index has a wide range around $-0.5$ to 0.2. Most galaxies sit in a locus around $C$-index = 0. There are a fraction of low $C$-index galaxies that stand out with $C$-index $\sim -0.2$. This shows that when progressing to higher halo mass environments the tail of galaxies with concentrated star-formation increases for high-$M_{\star}$ galaxies in 10$^{9.5-11.5}M_{\odot}$ $M_{\star}$ bin.

A Kolmogorov$-$Smirnov (K$-$S) test is applied on the $C$-index distribution in different environments for both the 10$^{9.5-11.5}M_{\odot}$ mass bin and the 10$^{7.5-9.5}M_{\odot}$ mass bin. A $p$-value below 0.05 allows us to reject the null hypothesis that the two distributions are the same. For the 10$^{9.5-11.5}M_{\odot}$ mass bin, the K-S test results for the four halo mass intervals are shown in Table \ref{tab:kstest}. The difference between ungrouped galaxies and those in low-mass groups is not significant ($p$-value = 0.397), but the difference between ungrouped galaxies and galaxies in high-mass groups or clusters is more significant ($p$-value = 0.030 and 1.025$\times 10^{-5}$, respectively). Similarly, for the low-mass groups, the difference between low-mass groups and high-mass group is less significant ($p$-value = 0.199) than with clusters ($p$-value = 0.003). These results show the $C$-index distribution in clusters and high-mass groups is not the same as in low density environments which supports the low $C$-index tail we find in clusters in Fig.~\ref{fig:massbin_agn}. We also carry out the K-S test without c\_L/A galaxies. This does not change the $p$-value significantly which confirms that adding c\_L/A does not bias our results.

The $C$-index in the 10$^{7.5-9.5}M_{\odot}$ mass bin shows a similar distribution for ungrouped galaxies and galaxies in low-mass groups ($p$-value = 0.38), and comparing low-mass groups and high-mass groups ($p$-value = 0.18). The ungrouped and high-mass group has a larger difference ($p$-value = 0.01). The larger $p$-values compared with high $M_{\star}$ galaxies suggest that galaxies with low $M_{\star}$ are affected by the environments.

To further quantify the distribution of regular galaxies and galaxies with concentrated \update{star-formation} (SF-concentrated galaxies) in different halo masses, our \update{Full\_sam} is separated into two intervals: $C$-index $\ge$ $-0.2$ and $C$-index $<$ $-0.2$. \update{We fit a Gaussian distribution based on the $C$-index of ungrouped galaxies}, resulting \update{in a mean, $\mu$} = 0.004 \update{and a standard deviation, $\sigma$} = 0.065. Therefore, we choose $-0.2$ as a valid cut off at 3 $\sigma$. The uncertainty on the fraction of SF-concentrated galaxies is calculated using the binomial distribution. For galaxies in the 10$^{9.5-11.5}M_{\odot}$ mass range, there are 9$\pm$2 per cent SF-concentrated galaxies in ungrouped regions, 8$\pm$3 per cent in low halo mass groups, 19$\pm$4 per cent in high halo mass groups and 29$\pm$4 per cent in clusters. For galaxies in the 10$^{7.5-9.5}M_{\odot}$ mass range, the fraction of SF-concentrated galaxies is 7$\pm$2 per cent for ungrouped regions, 9$\pm$2 per cent for low-mass groups and 24$\pm$7 per cent for high-mass groups. We also test the fraction of SF-concentrated galaxies using a $C$-index cut on $-0.1$ and $-0.3$, the trend remains the same that higher halo mass galaxies for both low and high $M_{\star}$ galaxies tend to have larger fractions of SF-concentrated galaxies.

\begin{figure*}
	\includegraphics[width=\textwidth]{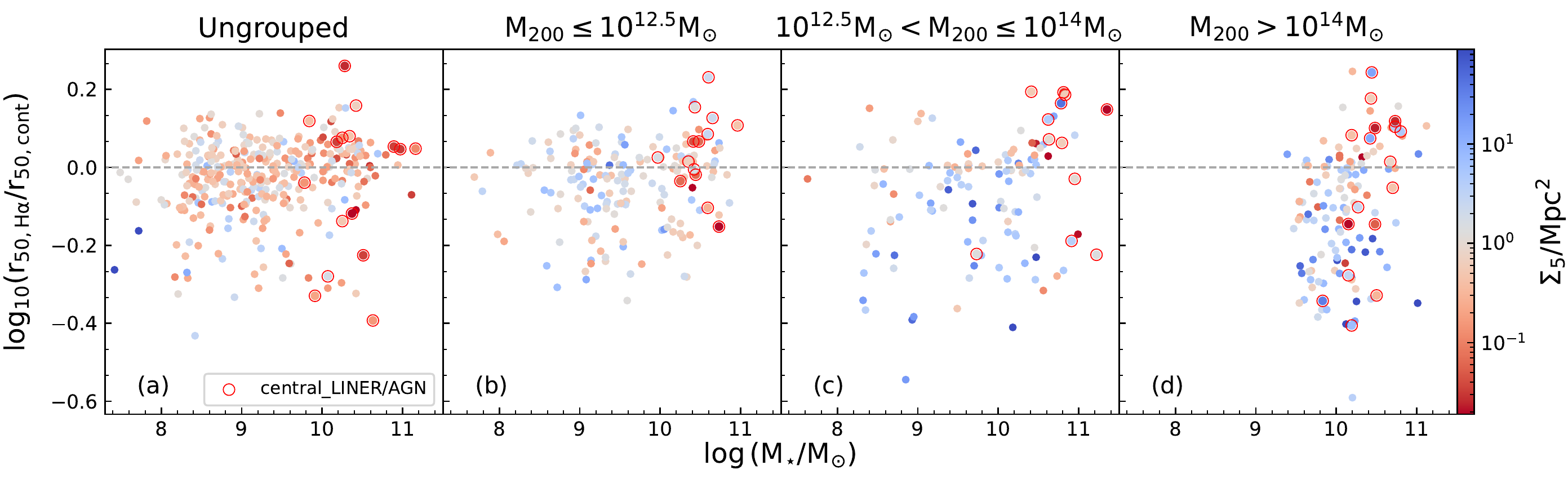}
    \caption{The ratio of $C$-index as a function of $M_{\star}$ in four halo mass intervals: ungrouped galaxies, low-mass groups, high-mass groups and clusters colour-coded by their surface density $\Sigma_{5}$/Mpc$^{2}$. The central$-$LINER/AGN galaxies with AGN correction are in red circles. Galaxies with higher $\Sigma_{5}$/Mpc$^{2}$ tend to have lower $C$-index.  }
    \label{fig:CvsMcolor5th}
\end{figure*}

\begin{figure}
	\includegraphics[width=7cm]{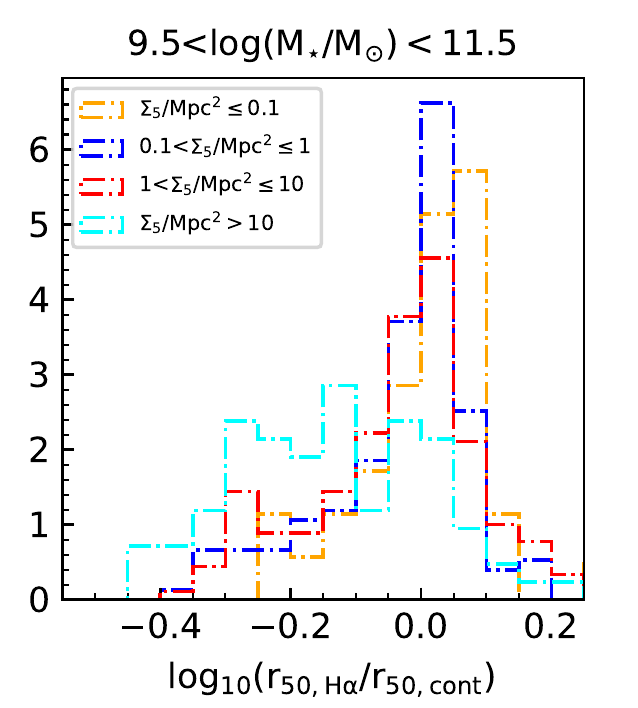}
    \caption{Histogram of $C$-index in different local environmental densities in \update{the 10$^{9.5-11.5}$ $M_{\odot}$ $M_{\star}$ bin in the Full\_sam}. $\Sigma_{5}$/Mpc$^{2}$ between 0-0.1 is shown in yellow, 0.1-1 in blue, 1-10 in red and >10 in cyan. Galaxies with $\Sigma_{5}$ > 10 Mpc$^{-2}$ have a more pronounced tail to low $C$-index. }
    \label{fig:histo_5th}
\end{figure}

\begin{table}
	\centering
	\caption{Results of the K-S test of the $C$-index in the Full\_sam within four local environmental density bins for galaxies with $M_{\star}$ of 10$^{7.5-11.5}$ $M_{\odot}$. \update{The statistic value ($D$) and probability ($p$) values are shown in the table.} We mark significant $p$-values in bold.}
	\label{tab:kstest_5th}
	\begin{tabular}{cccc} 
		\hline
		K$-$S test &  $\Sigma_{5}$/Mpc$^{2}$  &  $\Sigma_{5}$/Mpc$^{2}$ & $\Sigma_{5}$/Mpc$^{2}$ \\
		$ $ & 0.1 - 1 & 1-10 & >10\\
		\hline
		$\Sigma_{5}$/Mpc$^{2}$ & $D$=0.121, & $D$=0.165, & $D$=0.531, \\
		<0.1  &  $p$=0.608 & $p$=0.249 & $p$=\textbf{7.154e-08}\\
		\hline
		$\Sigma_{5}$/Mpc$^{2}$  & $ $ & $D$=0.110, & $D$=0.441,\\
		0.1 - 1 &$ $& $p$=\textbf{0.046} & $p$=\textbf{5.129e-13}\\
		\hline
		$\Sigma_{5}$/Mpc$^{2}$ & $ $ & $ $ & $D$=0.397\\
		1-10 &$ $& $ $&$p$=\textbf{3.018e-10}\\
		\hline
	\end{tabular}
\end{table}

\subsection{The spatial star-formation distribution versus local surface density}

For local environmental densities, we select four intervals in $\Sigma_{5}$: 0-0.1, 0.1-1, 1-10 and >10 Mpc$^{-2}$. In Fig.~\ref{fig:CvsMcolor5th}, we show $C$-index as a function of $M_{\star}$, in our usual halo mass intervals, but colour-coded by log($\Sigma_{5}$/Mpc$^{2}$). There is a wide range of $\Sigma_{5}$ in the 4 halo mass ranges. The histograms of $C$-index in 10$^{9.5-11.5}M_{\odot}$ mass bin in different $\Sigma_{5}$ intervals is shown in Fig.~\ref{fig:histo_5th}. There is a clear difference between $\Sigma_{5}$ < 10 Mpc$^{-2}$ and $\Sigma_{5}$ > 10 Mpc$^{-2}$ galaxies with $\Sigma_{5}$ > 10 Mpc$^{-2}$ having a more pronounced tail to low $C$-index. Concentrated SF galaxies tend to live in higher local densities. To quantify the difference, a K-S test is applied on the $C$-index in 4 $\Sigma_{5}$ intervals and the results are shown in Table \ref{tab:kstest_5th}. We find that the $C$-index distribution for galaxies with $\Sigma_{5}$>10 Mpc$^{-2}$ is significantly different to that of galaxies with lower $\Sigma_{5}$ with $p-$value < 0.05. However, there is no significant difference between the $C$-indices in the lower $\Sigma_{5}$ intervals.  

\begin{figure}
	\includegraphics[width=\columnwidth]{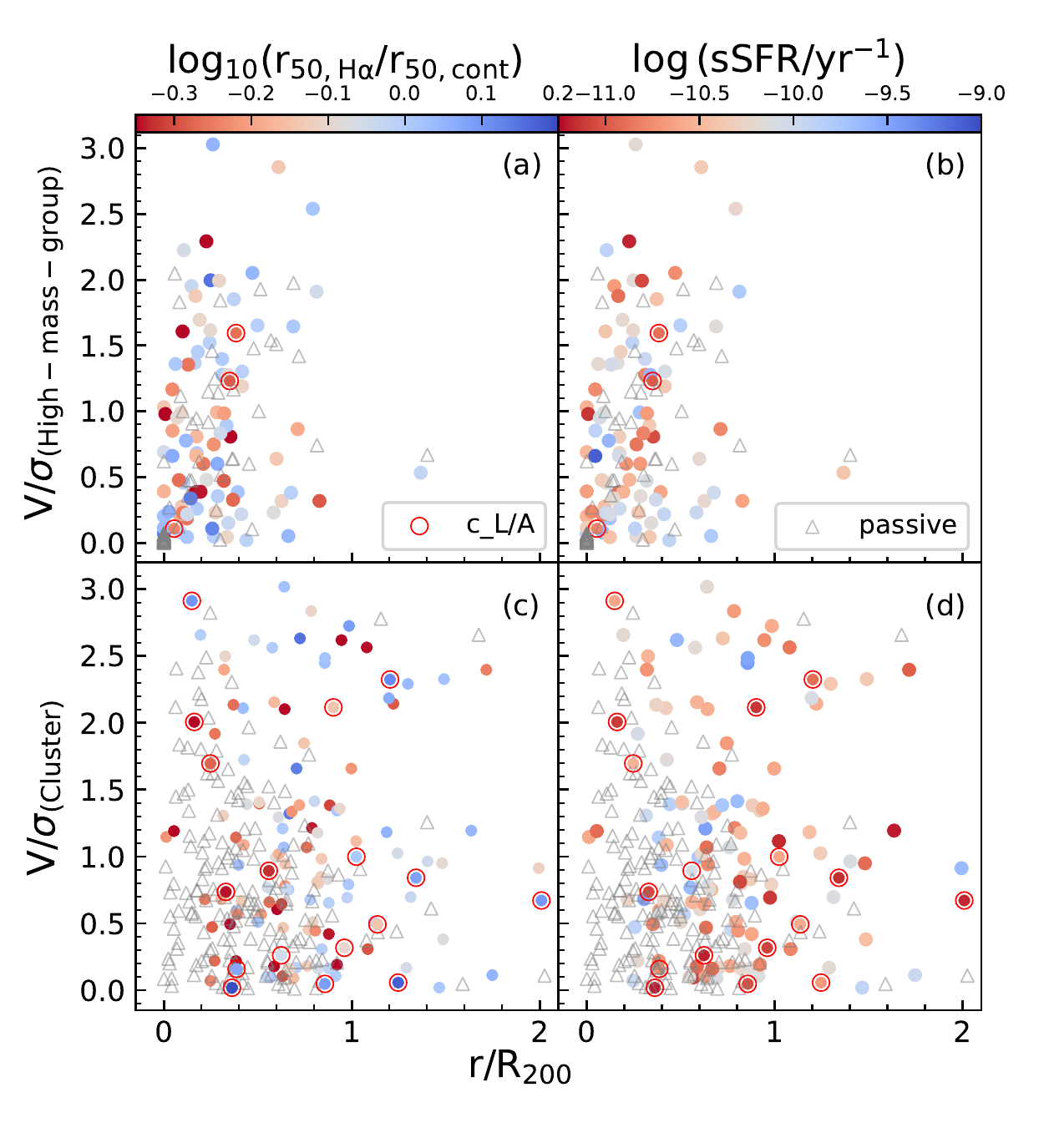}
    \caption{Projected phase-space diagram for satellite galaxies in high-mass groups and clusters. The passive galaxies are shown by grey triangles. The horizontal axis is the projected distance of an individual galaxy from the centre of the halo \update{normalized by the radius of the halo $R_{200}$}. The vertical axis is the velocity of an individual galaxy relative to the systemic velocity of the halo, normalized by the velocity dispersion of the halo. Panels a and c are colour-coded by $C$-index and panels b and d are colour-coded by sSFR. The c\_L/A galaxies are highlighted by red circles. }
    \label{fig:phasespace}
\end{figure}

\subsection{Projected phase-space for group and cluster galaxies}

\update{In the phase-space diagram, galaxies closer to the centre (within the virial radius), are more likely to undergo ram-pressure stripping \citep[][]{Jaffe2015}. }The phase-space diagram for \update{Full\_sam} colour-coded by $C$-index (left column) and sSFR (right column) for galaxies is shown in Fig.~\ref{fig:phasespace}. \update{Here we only plot the satellites as we expect these to be most influenced by environments}. The passive galaxies with EW$_\mathrm{H\alpha} \le$ 1\AA\ are shown as grey triangles to show the complete distribution. We do not consider low-mass groups in the phase-space diagram as there are fewer group members, and their groups have large errors on their velocity dispersions. To check any distribution difference between SF-concentrated ($C$-index < $-0.2$) and regular galaxies in the phase-space diagram, we apply a 2D K-S test. In the high-mass groups (panels a, b), the $p$-value from the 2D K-S test is equal to 0.31, there is not a significant difference between SF-concentrated galaxies and regular galaxies in the phase-space diagram. Also, the passive galaxy distribution does not have a significant difference from the SF galaxies. In clusters (panels c, d), only passive galaxies are seen near the cluster centre (within 0.2 $R_{200}$). For SF galaxies, within 0.5 $R_{200}$, 50 per cent of the galaxies have a centrally concentrated SF. The $p$-value from a 2D K-S test on $C$-index cut ($C$-index=$-0.2$) equals 0.03 for clusters, meaning the $C$-index distributions are different. Galaxies with concentrated \update{star-formation} are expected to be located nearer the centre of the cluster as those galaxies may be more impacted by quenching processes such as RPS.

The phase-space diagram \update{indicates where the quenched galaxies currently are} but this is not sufficient to see how fast the quenching happened. \update{For example, there are studies suggesting galaxies may have already started quenching or already finished quenching before they fall into the current host \citep[e.g.][]{Mihos2004, Oh_2018}. Results from numerical simulations also show good agreements with observations on environmental effects in the outskirts of clusters \citep[e.g.][]{Ayromlou2019, Ayromlou2021, Coenda2021}. Along with pre-processing, galaxy quenching in clusters is also effected by  their orbit, for example their closest pericentric approach \citep[e.g.][]{Arthur2019, DiCintio2021}.} In the next section, we further investigate the SF-concentrated galaxies by adding information about the stellar population ages, derived using both indices (such as $D_\mathrm{n}$4000, $H\delta_\mathrm{A}$) and full spectral fitting.

\begin{figure*}
	\includegraphics[width=\textwidth]{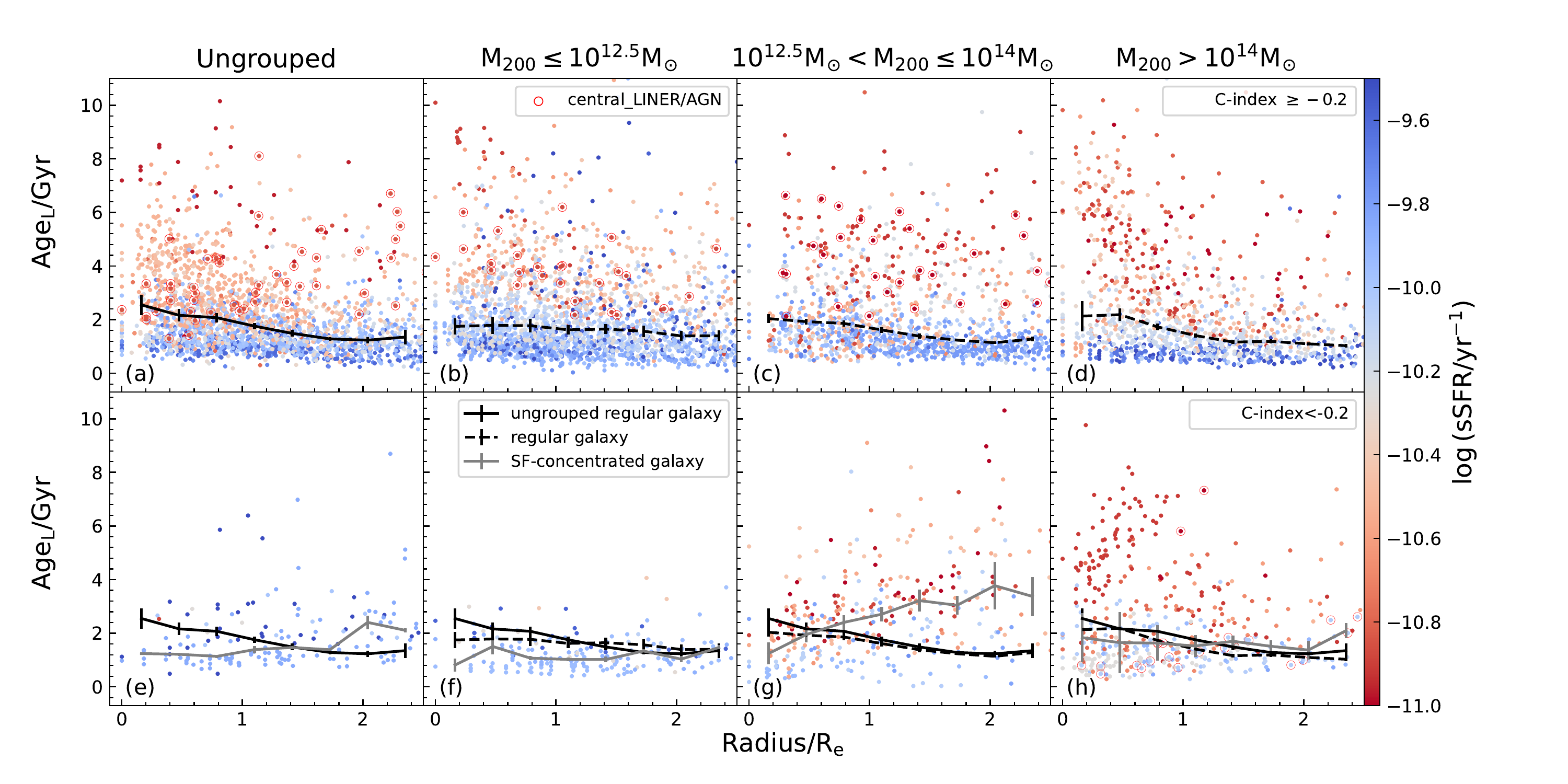}
    \caption{$Age_\mathrm{L}$ as a function of galactocentric radius in four halo mass intervals colour-coded by total sSFR. The top row shows regular SF galaxies with $C$-index $\ge$ $-0.2$. The bottom row shows SF galaxies with $C$-index $<$ $-0.2$. In each panel, we create 8 bins from 0 - 2.5 $R_\mathrm{e}$ and plot the median $Age_\mathrm{L}$ of each bin with bootstrapping uncertainties. The black lines are from regular galaxies and the grey lines are from the concentrated galaxies. The solid black line is the median from the ungrouped regular galaxies, shown in panels e, f, g and h. The dashed black lines are the medians from the group/cluster regular galaxies and are also shown in the concentrated galaxies panels in the corresponding environment. Regular galaxies in all environments show older centres with younger discs. SF-concentrated galaxies in high-mass groups show older discs than cluster galaxies.  }
    \label{fig:AGEvsRe_LW_SFR}
\end{figure*}

\begin{figure*}
	\includegraphics[width=\textwidth]{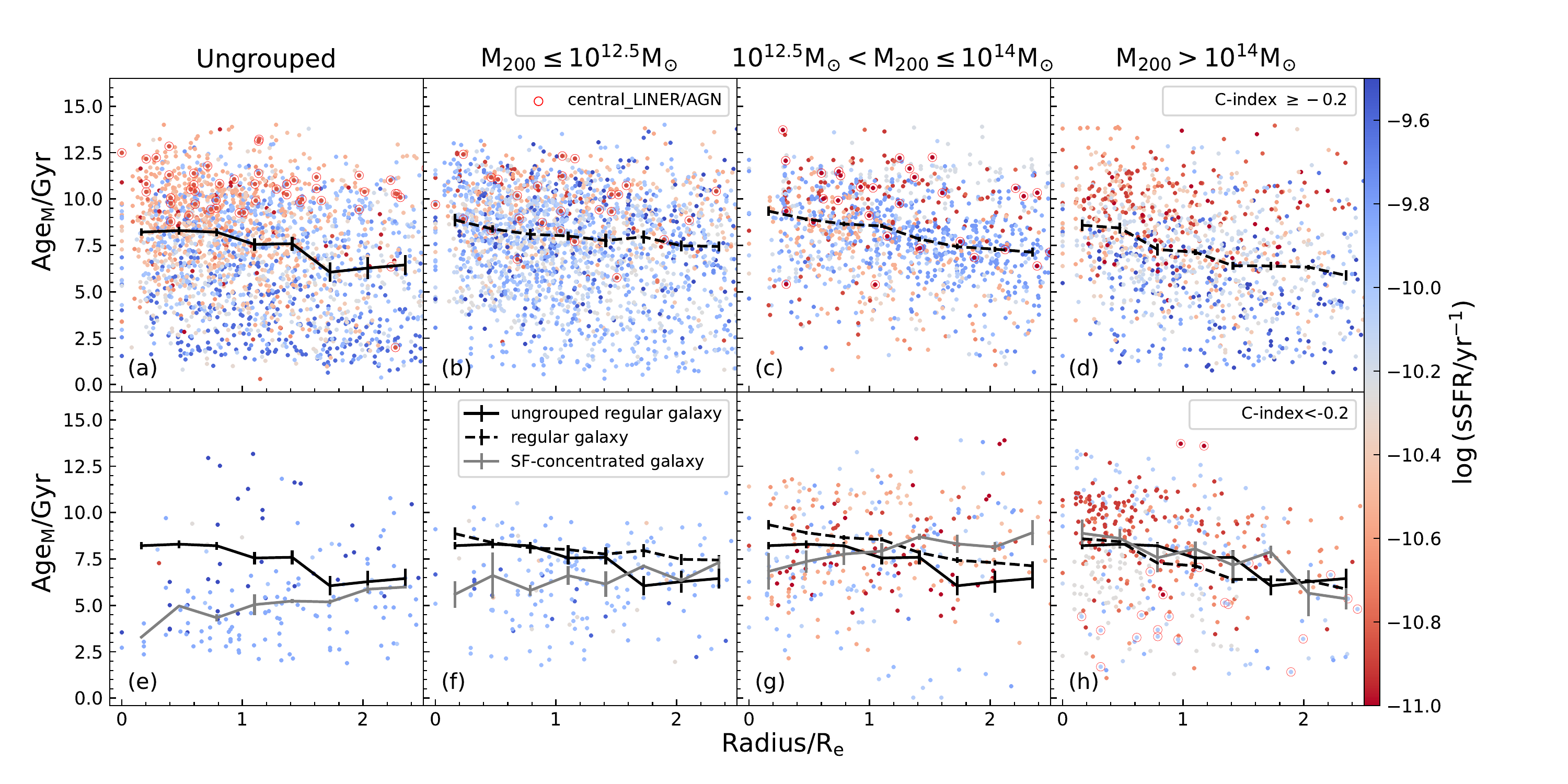}
    \caption{$Age_\mathrm{M}$ as a function of galactocentric radius in four halo mass intervals colour-coded by total sSFR. In each panel, we create 8 bins from 0 - 2.5 $R_\mathrm{e}$ and plot the median $Age_\mathrm{M}$ of each bin with bootstrapping uncertainties. The lines and error-bars are in the same formats as Fig.~\ref{fig:AGEvsRe_LW_SFR}. The $Age_\mathrm{M}$ radial profile is broadly consistent with $Age_\mathrm{L}$ radial profile. High-mass group SF-concentrated galaxies have older discs than ungrouped regular galaxies. This age difference disappears in clusters. }
    \label{fig:AGEvsRe_MW_SFR}
\end{figure*}

\begin{figure*}
	\includegraphics[width=\textwidth]{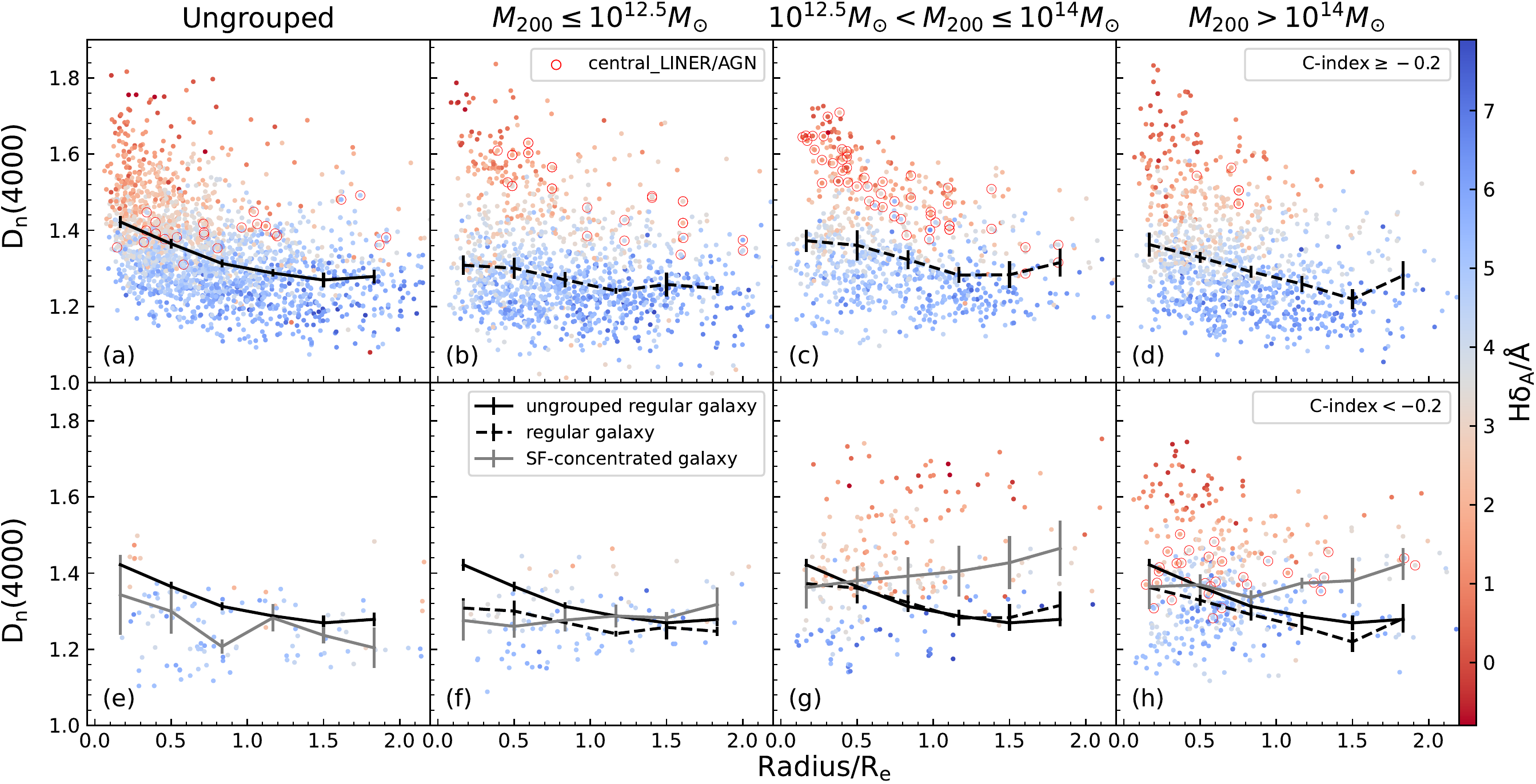}
    \caption{$D_\mathrm{n}$4000 as a function of galactocentric radius in four halo mass intervals colour-coded by $H\delta_\mathrm{A}$. In each panel, we create 6 bins from 0 - 2 $R_\mathrm{e}$ and plot the median $D_\mathrm{n}$4000 value of each bin with bootstrapping uncertainties. The lines and error-bars are in the same formats as Fig.~\ref{fig:AGEvsRe_LW_SFR}. SF-concentrated galaxies in high-mass groups and clusters have higher $D_\mathrm{n}$4000 indices than regular galaxies. }
    \label{fig:DnvsRe}
\end{figure*}

\begin{figure*}
	\includegraphics[width=\textwidth]{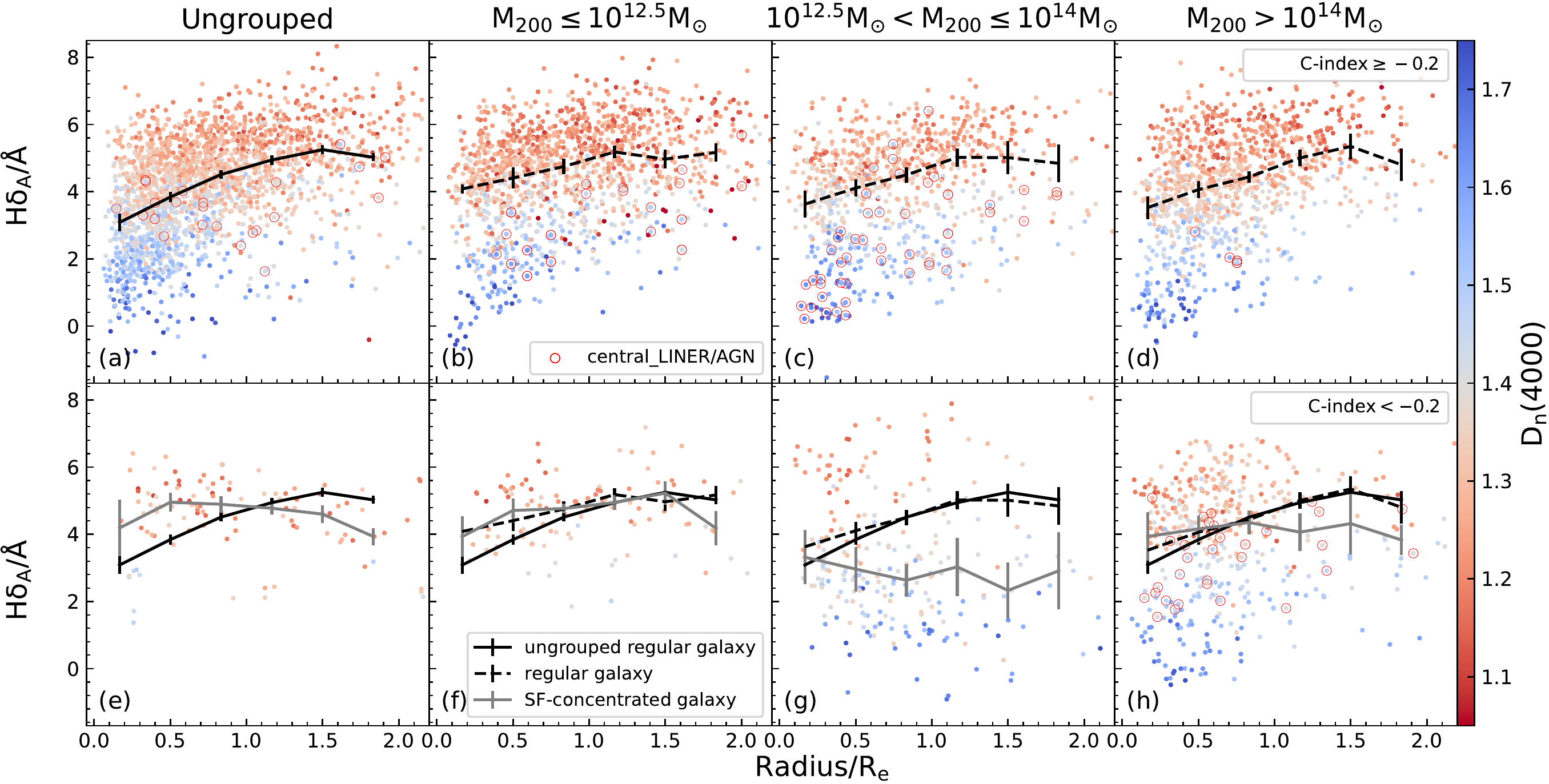}
    \caption{$H\delta_\mathrm{A}$ as a function of galactocentric radius in four halo mass intervals colour-coded by $D_\mathrm{n}$4000. The top row shows regular galaxies while the bottom row shows concentrated galaxies. The lines and error-bars are in the same formats as Fig.~\ref{fig:AGEvsRe_LW_SFR}. SF-concentrated galaxies in high-mass groups and clusters have lower $H\delta_\mathrm{A}$ indices than regular galaxies. }
    \label{fig:HdevsRe}
\end{figure*}

\section{Results: \update{radial age gradients}}
\label{sec:result_quenching} 
The light-weighted age ($Age_\mathrm{L}$), mass-weighted age ($Age_\mathrm{M}$), $D_\mathrm{n}$4000 and $H\delta_\mathrm{A}$, are calculated from the stellar population fits (more details in Section~\ref{sec:dn4000}) and used here to better understand the quenching time-scale. The \update{442} galaxies with 10$^{9.5-11.5}M_{\odot}$ mass are included in this section to match the $M_{\star}$ range in the clusters. After we apply $D_\mathrm{n}$4000 and $H\delta_\mathrm{A}$ uncertainty limits ($H\delta_\mathrm{A}$ error < 1 \AA $ $ and $D_\mathrm{n}$4000 index error < 0.1), the total number of galaxies studied here is 350 (92 galaxies removed).

\begin{table*}
	\centering
	\caption{Median $Age_\mathrm{L}$, $Age_\mathrm{M}$, $D_\mathrm{n}$4000 and $H\delta_\mathrm{A}$measurements with uncertainty for 0-1 $R_\mathrm{e}$ and 1-2 $R_\mathrm{e}$ for galaxies with $M_{\star}$ larger than 10$^{9.5}$ $M_{\odot}$. Regular galaxies show older ages in centres with younger populations in the outer region. SF-concentrated galaxies in ungrouped regions and low-mass groups show younger centres than the outer regions. SF-concentrated galaxies in high-mass group show significant older discs while this signature disappears in clusters. }
	\label{tab:dn_table}
	\begin{tabular}{ccccc} 
		\hline
		Stellar index & Ungrouped galaxies & Group galaxies & Group galaxies& Cluster galaxies \\
		$ $ & $ $ & (halo mass${\le} 10^{12.5} M_{\odot}$)&(halo mass in $ 10^{12.5-14} M_{\odot}$)&(halo mass${>} 10^{14} M_{\odot}$)\\
		\hline
		$ $ & 0-1$R_\mathrm{e}$ 1-2$R_\mathrm{e}$ & 0-1$R_\mathrm{e}$ 1-2$R_\mathrm{e}$ &0-1$R_\mathrm{e}$ 1-2$R_\mathrm{e}$ &0-1$R_\mathrm{e}$ 1-2$R_\mathrm{e}$ \\
		\hline
		$Age_\mathrm{L}$/Gyr (Reg)& 2.09$\pm$0.18$\ $1.42$\pm$0.11 & 1.74$\pm$0.27$\ $1.57$\pm$0.22 & 1.86$\pm$0.10$\ $1.30$\pm$0.10 & 1.93$\pm$0.14$\ $1.15$\pm$0.06\\ (Concentrated) & 1.20$\pm$0.07$\ $1.61$\pm$0.10 & 1.12$\pm$0.14$\ $1.06$\pm$0.03 & 1.95$\pm$0.18$\ $3.25$\pm$0.37 & 1.66$\pm$0.75$\ $1.61$\pm$0.17 \\ 
		\hline
		$Age_\mathrm{M}$/Gyr (Reg)& 8.19$\pm$0.20$\ $6.93$\pm$0.43 & 8.33$\pm$0.21$\ $7.69$\pm$0.26 & 8.76$\pm$0.12$\ $7.72$\pm$0.12 & 8.00$\pm$0.26$\ $6.42$\pm$0.23\\ (Concentrated) & 4.49$\pm$0.21$\ $5.45$\pm$0.19 & 6.50$\pm$0.62$\ $6.58$\pm$0.18 & 7.65$\pm$0.53$\ $8.26$\pm$0.35 & 8.13$\pm$0.62$\ $7.33$\pm$0.42 \\
		\hline
		$D_\mathrm{n}$4000(Reg) & 1.36$\pm$0.01$\ $1.28$\pm$0.01 & 1.29$\pm$0.02$\ $1.25$\pm$0.01 & 1.35$\pm$0.02$\ $1.28$\pm$0.02 & 1.33$\pm$0.02$\ $1.25$\pm$0.02\\ (Concentrated) & 1.28$\pm$0.03$\ $1.25$\pm$0.03 & 1.27$\pm$0.03$\ $1.28$\pm$0.01 & 1.38$\pm$0.04$\ $1.41$\pm$0.05 & 1.36$\pm$0.03$\ $1.39$\pm$0.02 \\
		\hline
		$H\delta_\mathrm{A}$(Reg) & 3.89$\pm$0.14$\ $5.04$\pm$0.11 & 4.42$\pm$0.19$\ $5.14$\pm$0.20 & 4.11$\pm$0.26$\ $4.99$\pm$0.35 & 4.09$\pm$0.21$\ $5.07$\pm$0.25\\ (Concentrated) & 4.87$\pm$0.25$\ $4.58$\pm$0.15 & 4.76$\pm$0.28$\ $4.95$\pm$0.24 & 3.07$\pm$0.58$\ $2.91$\pm$0.62 & 4.15$\pm$0.33$\ $4.01$\pm$0.50 \\
		\hline
	\end{tabular}
\end{table*}

\begin{table*}
	\centering
	\caption{Assuming that prior to entering a group/cluster the average age profile is similar to that of the ungrouped galaxies, the regular ungrouped galaxy profiles can be compared to the SF-concentrated galaxies. We list the $Age_\mathrm{L}$, $Age_\mathrm{M}$, $D_\mathrm{n}$4000 and $H\delta_\mathrm{A}$ indices for 0-1 $R_\mathrm{e}$ and 1-2 $R_\mathrm{e}$ subtracting the values of ungrouped regular galaxies (panel a in Fig~\ref{fig:AGEvsRe_LW_SFR}, \ref{fig:AGEvsRe_MW_SFR}, \ref{fig:DnvsRe}, \ref{fig:HdevsRe}) for galaxies with $M_{\star}$ larger than 10$^{9.5}$ $M_{\odot}$. We mark the indices > 2$\sigma$ in bold. The index differences are the greatest in high-mass groups for SF-concentrated galaxies. }
	\label{tab:dn_diff_table}
	\begin{tabular}{ccccc} 
		\hline
		Stellar index & Ungrouped galaxies & Group galaxies & Group galaxies& Cluster galaxies \\
		difference & $ $ & (halo mass${\le} 10^{12.5} M_{\odot}$)&(halo mass in $ 10^{12.5-14} M_{\odot}$)&(halo mass${>} 10^{14} M_{\odot}$)\\
		\hline
		$ $ & 0-1$R_\mathrm{e}$ 1-2$R_\mathrm{e}$ & 0-1$R_\mathrm{e}$ 1-2$R_\mathrm{e}$ &0-1$R_\mathrm{e}$ 1-2$R_\mathrm{e}$ &0-1$R_\mathrm{e}$ 1-2$R_\mathrm{e}$ \\
		\hline
		$Age_\mathrm{L}$/Gyr (Reg) & $\ $ & -0.35$\pm$0.32$\ $0.15$\pm$0.25 & -0.23$\pm$0.20$\ $-0.12$\pm$0.15 & -0.16$\pm$0.22$\ $-0.26$\pm$0.13\\ (Concentrated) & \textbf{-0.89$\pm$0.19}$\ $0.19$\pm$0.15 & \textbf{-0.97$\pm$0.23}$\ $\textbf{-0.36$\pm$0.12} & -0.14$\pm$0.25$\ $\textbf{1.83$\pm$0.38} & -0.43$\pm$0.77$\ $0.19$\pm$0.21 \\
		\hline
		$Age_\mathrm{M}$/Gyr (Reg)& $\ $ & 0.14$\pm$0.29$\ $0.76$\pm$0.50 & \textbf{0.57$\pm$0.23}$\ $0.79$\pm$0.45 & -0.19$\pm$0.33$\ $-0.50$\pm$0.49\\ (Concentrated) & \textbf{-3.70$\pm$0.29}$\ $\textbf{-1.48$\pm$0.48} & \textbf{-1.69$\pm$0.65}$\ $-0.34$\pm$0.47 & -0.53$\pm$0.57$\ $\textbf{1.34$\pm$0.56} & -0.06$\pm$0.65$\ $0.40$\pm$0.61 \\ 
		\hline
		$D_\mathrm{n}$4000(Reg) &$\ $ & -0.07$\pm$0.02$\ $-0.03$\pm$0.02 & -0.01$\pm$0.03$\ $0.00$\pm$0.02 & -0.03$\pm$0.02$\ $-0.03$\pm$0.02\\ (Concentrated) & \textbf{-0.08$\pm$0.03}$\ $-0.03$\pm$0.03 & \textbf{-0.09$\pm$0.03}$\ $0.00$\pm$0.02 & 0.02$\pm$0.04$\ $\textbf{0.13$\pm$0.05} & 0.00$\pm$0.03$\ $\textbf{0.11$\pm$0.03} \\ 
		\hline
		$H\delta_\mathrm{A}$(Reg) & $\ $ & \textbf{0.53$\pm$0.23}$\ $0.10$\pm$0.23 & 0.22$\pm$0.30$\ $-0.06$\pm$0.36 & 0.20$\pm$0.26$\ $0.03$\pm$0.27\\ (Concentrated) & \textbf{0.98$\pm$0.29}$\ $\textbf{-0.46$\pm$0.18} & \textbf{0.87$\pm$0.32}$\ $-0.10$\pm$0.26 & -0.83$\pm$0.60$\ $\textbf{-2.14$\pm$0.63} & 0.25$\pm$0.36$\ $-1.03$\pm$0.52 \\ 
		\hline
		
	\end{tabular}
\end{table*}

\subsection{Age radial profile}
\label{sec:age radial profile} 

The stellar population age calculated by full spectral fitting is described in Section~\ref{sec:dn4000}. $Age_\mathrm{L}$ and $Age_\mathrm{M}$ for galaxies with $M_{\star}$ of 10$^{9.5-11.5}$ $M_{\odot}$ as a function of galactocentric radius in different $C$-index bins are shown in Fig.~\ref{fig:AGEvsRe_LW_SFR} and Fig.~\ref{fig:AGEvsRe_MW_SFR}. The x axis is the elliptical radius in units of$ $ $R_\mathrm{e}$. From left to right, the panels show our four bins in halo mass. Each dot represents an individual sector bin in a SAMI galaxy, colour-coded by the integrated sSFR for each galaxy. The sample is separated at $C$-index = $-0.2$ (regular galaxies with $C$-index $\ge$ $-0.2$ and SF-concentrated galaxies with $C$-index $<$ $-0.2$) to see the difference as a function of star-formation concentration. The medians in 8 bins between 0-2.5 $R_\mathrm{e}$ are shown by the lines with error-bars. The uncertainties on the medians are calculated using bootstrapping statistics by galaxies to better capture the intrinsic variation between galaxies rather than just per radius bin/sector. The solid black line is for regular ungrouped galaxies (in panel a) and it is plotted as a reference line in panels e-h. The dashed black lines are the medians for group/cluster regular galaxies. To see the difference between regular and concentrated galaxies, the dashed black lines are plotted in the corresponding environment in panels f-h. The grey lines are for the SF-concentrated galaxies in panels e-h. 

For both $Age_\mathrm{L}$ and $Age_\mathrm{M}$, age is decreasing with an increasing radius for regular galaxies (panels a-d) which is consistent with the well-known `inside-out' galaxy growth \citep[e.g.][]{Perez2013, Gonzalez2015}. $Age_\mathrm{L}$ is weighted more towards the most recent starbursts while $Age_\mathrm{M}$ provides a range averaged across the entire stellar population. So that we also find the age radial profiles are steeper in $Age_\mathrm{M}$ than $Age_\mathrm{L}$. \update{We also see sSFR is higher for younger $Age_\mathrm{L}$; in $Age_\mathrm{M}$, sSFR is more mixed.} \update{The `inside-out' growth can happen when gas in the centre with low angular momentum cools down and forms stars on shorter timescales than gas of high angular momentum in the disc \citep[e.g.][]{Larson1978, Gogarten2010, Frankel2019, Sacchi2019}. Along with observational results, the `inside-out' growth is also supported by hydrodynamical simulations \citep[e.g.][]{Somerville2008, AvilaReese2018}.}

For SF-concentrated galaxies in ungrouped and low-mass groups (Fig.~\ref{fig:AGEvsRe_MW_SFR},~\ref{fig:AGEvsRe_LW_SFR}e, f), the inner regions have lower values than ungrouped regular galaxies and they have similar ages in the outskirts. As we are selecting SF-concentrated galaxies by $C$-index, the younger ages may be caused by central \update{star-formation} enhancements. For SF-concentrated galaxies in high-mass groups (Fig.~\ref{fig:AGEvsRe_MW_SFR},~\ref{fig:AGEvsRe_LW_SFR}g), we find a distinct signature that although the inner centre ($\sim$0.2 $R_\mathrm{e}$) is lower than ungrouped regular galaxies, the ages in the outer region is significantly older. We also notice the SF-concentrated galaxies in high-mass groups show greater $Age_\mathrm{L}$ difference in the outer region than $Age_\mathrm{M}$. The older ages at large radii, pointing to older discs (relative to ungrouped regular galaxies). Interestingly, this older discs signature does not show up in clusters (Fig.~\ref{fig:AGEvsRe_MW_SFR},~\ref{fig:AGEvsRe_LW_SFR}h). SF-concentrated galaxies in clusters show declining age radial profiles with radius as regular galaxies. And there is no age difference for both $Age_\mathrm{L}$ and $Age_\mathrm{M}$ at a large radius compared to ungrouped regular galaxies.  

To quantify the difference between the age measurements for galaxy centres and outskirts, we calculate the median value for the radial intervals 0 - 1 $R_\mathrm{e}$ and 1 - 2 $R_\mathrm{e}$ for each panel and we present them in Table~\ref{tab:dn_table}. We find a median outer $Age_\mathrm{L}$ ($Age_\mathrm{M}$) of 3.35$\pm$0.37 Gyr (8.26$\pm$0.35 Gyr) for SF-concentrated galaxies in high-mass groups, 1.61$\pm$0.17 Gyr (7.33$\pm$0.41 Gyr) for SF-concentrated galaxies in clusters and 1.61$\pm$0.10 Gyr (6.93$\pm$0.43 Gyr) for ungrouped regular galaxies. Assuming that prior to entering a group/cluster the average age profile is similar to that of the ungrouped galaxies, the regular ungrouped galaxy profiles can be compared to the SF-concentrated galaxies. Compared with ungrouped regular galaxies, the age differences with SF-concentrated galaxies are given in Table~\ref{tab:dn_diff_table}. The age difference is greatest for high-mass groups (1.83$\pm$0.38 Gyr for $Age_\mathrm{L}$; 1.34$\pm$0.56 Gyr for $Age_\mathrm{M}$). The age difference in clusters are not significant (0.19$\pm$0.21 Gyr for $Age_\mathrm{L}$; 0.40$\pm$0.61 Gyr for $Age_\mathrm{M}$). We find that SF-concentrated galaxies in high-mass groups have relatively older discs than those in clusters.

\subsection{$D_\mathrm{n}$4000 index radial profile}

To further examine the age trends, we repeat the analysis with $D_\mathrm{n}$4000 index and $H\delta_\mathrm{A}$ index. These spectral features are direct measurements from SAMI cubes and are age sensitive, so those indices can be used to test whether we have the same qualitative trends as the age estimates from full spectral fitting. We show the $D_\mathrm{n}$4000 index versus radius for galaxies of $M_{\star}$ of 10$^{9.5-11.5}$ $M_{\odot}$ for two bins in $C$-index in Fig.~\ref{fig:DnvsRe}. The plot is similar to Fig.~\ref{fig:AGEvsRe_LW_SFR}. The medians in each panel with bootstrapping uncertainties are shown by the lines with error-bars. The bootstrapping uncertainty is relatively large in Fig.~\ref{fig:DnvsRe}f and Fig.~\ref{fig:HdevsRe}f because there are only 8 galaxies in those subplots.

When comparing the medians in the $D_\mathrm{n}$4000 profiles of regular galaxies in different environments, they are similar with the standard deviations of around 0.03. Regular galaxies (Fig.~\ref{fig:DnvsRe} a-d) all show a general decreasing $D_\mathrm{n}$4000 with increasing radius. The $D_\mathrm{n}$4000 gradient for regular galaxies is consistent with the age radial profiles, where central stellar populations are older. In panels e and f (ungrouped and low-mass groups) the SF-concentrated galaxies appear to have lower $D_\mathrm{n}$4000 at a small radius than regular galaxies, as might be expected given they have concentrated star-formation. However, they appear to have $D_\mathrm{n}$4000 consistent with regular galaxies at a larger radius. This may point toward these galaxies being dominated by central \update{star-formation} enhancements, but we caution that the number of galaxies in these samples is small (8 galaxies in both panels e and f). Both high-mass groups and clusters (Fig.~\ref{fig:DnvsRe} g, h) show slightly lower $D_\mathrm{n}$4000 in the centres, but higher $D_\mathrm{n}$4000 at large radius, pointing to older discs (relative to regular galaxies). For high-mass group environments (panel g), in the centre, $D_\mathrm{n}$4000 is slightly below the regular galaxies but at a larger radius, $D_\mathrm{n}$4000 is considerably higher. 

To quantify the difference of $D_\mathrm{n}$4000 index for the inner and outer parts of galaxies, the median values of 0 - 1 $R_\mathrm{e}$ and 1 - 2 $R_\mathrm{e}$ for each panel are calculated in Table~\ref{tab:dn_table}. The $D_\mathrm{n}$4000 index for the outer regions is 1.41$\pm$0.05 for SF-concentrated galaxies in the high-mass groups, 1.39$\pm$0.02 for SF-concentrated galaxies in cluster and 1.28$\pm$0.02 for ungrouped regular galaxies. The differences compared with ungrouped regular galaxy are in Table~\ref{tab:dn_diff_table}. The $D_\mathrm{n}$4000 difference in high-mass groups at 1 - 2 $R_\mathrm{e}$ is 0.13$\pm$0.05 while in clusters is 0.11$\pm$0.03. Galaxies in high-mass groups and clusters both show older discs than ungrouped regular galaxies. At large radius, although $D_\mathrm{n}$4000 indices seem to be larger for SF-concentrated galaxies in clusters, they are corresponding to ages considering uncertainties.

\subsection{$H\delta_\mathrm{A}$ index radial profile}

Along with $D_\mathrm{n}$4000, we also show $H\delta_\mathrm{A}$ measurements which correspond to more recent star-formation. The $H\delta_\mathrm{A}$ radial profiles in Fig.~\ref{fig:HdevsRe} for regular galaxies (panels a-d) all show an increasing $H\delta_\mathrm{A}$ with increasing radius. This is consistent scenario as the age radial profiles which shows an `inside-out' quenching process. Regular galaxies in different halo mass intervals (black solid lines and black dashed lines) do not have significant differences as the standard deviations of median values are around 0.25 \AA. In ungrouped and low-mass groups (panels e and f), the SF-concentrated galaxies tend to have higher $H\delta_\mathrm{A}$ at small radius than regular galaxies as might be expected given they are concentrated and may have recent star-formation. They tend to have slightly smaller $H\delta_\mathrm{A}$ at large radius compared to regular galaxies which may be because these galaxies are dominated by central \update{star-formation} enhancements. There is a larger difference between concentrated and regular galaxies in high-mass groups and clusters than low density environments. For SF-concentrated galaxies at large radius, $H\delta_\mathrm{A}$ is considerably lower in high-mass groups. 

The medians of 0 - 1 $R_\mathrm{e}$ and 1 - 2 $R_\mathrm{e}$ in each panel in Fig.~\ref{fig:HdevsRe} are given in Table~\ref{tab:dn_table}. The outer $H\delta_\mathrm{A}$ index is 2.91$\pm$0.62 \AA\ for SF-concentrated galaxies in the high-mass groups, 4.01$\pm$0.33 \AA\ for cluster SF-concentrated galaxies and 3.89$\pm$0.14 \AA\ for ungrouped regular galaxies. As $H\delta_\mathrm{A}$ is low at a large radius in high-mass groups, \update{there is }little evidence of a recently quenched population. The clusters have slightly higher $H\delta_\mathrm{A}$ than high-mass groups but less than regular galaxies which suggests that those populations have aged less. The median $H\delta_\mathrm{A}$ values are relatively low, suggesting recent star-formation. The index differences between the various populations and the ungrouped regular galaxies are given in Table~\ref{tab:dn_diff_table}. There is a clear difference between clusters and high-mass groups and the indices suggest relatively old populations in the outer discs for high-mass groups, meaning they quench the outer regions earlier which supports our age measurements.

\section{Discussion}\label{sec:discussion}

We aim to better understand the spatially-resolved environmental star-formation quenching processes by leveraging the comprehensive environment coverage of the SAMI Galaxy Survey. We combine spatial star-formation analysis with stellar-population properties, thus combining the `instantaneous' approach to quenching to a time-integrated analysis. With the final SAMI data release 3, the concentration index [$C$-index, log($r_\mathrm{50,H\alpha}/r_\mathrm{50,cont}$)] and SFR \update{are used} to test how spatial star-formation depends on environment. With these parameters, we are mainly selecting galaxies that currently show evidence of quenching. The stellar population radial profiles are used to \update{understand the relative rapidity of quenching in different environments.}

We find higher halo masses tend to have larger fractions of SF-concentrated galaxies than low halo masses. In some cases we find that SF-concentrated galaxies have older outskirts. Specifically, the age difference for the outer regions between SF-concentrated galaxies and regular galaxies is larger in high-mass groups (1.83$\pm$0.38 Gyr for $Age_\mathrm{L}$, 1.34$\pm$0.56 Gyr for $Age_\mathrm{M}$) than clusters ($0.19\pm$0.21 Gyr for $Age_\mathrm{L}$, 0.40$\pm$0.61 Gyr for $Age_\mathrm{M}$). The spatial extent of star-formation of low $M_{\star}$ galaxies ($10^{7.5-9.5} M_{\odot}$) also appears to be affected by the environments. 

Some papers, that did not find that the SF population depends on the environment, suggested the environmentally triggered transformations must occur either on a short timescale or preferentially at high redshift \citep[e.g.][]{Balogh_2004,Wijesinghe_2012}. While there are other works that show environmental impact on the SF population \citep[e.g.][]{vonder2010}. \update{There are studies in the Virgo cluster \citep[e.g.][]{Koopmann2004, Boselli2016} that show the SF population changes with increasing environmental density, likely reflecting the increasing impact of ram pressure.} With integral field data, we can see that although the passive fraction does change strongly with environment, there is a change in the SF population as well. The possible physical quenching processes in different halo mass intervals are discussed in the following sections.

\subsection{Quenching in clusters}

In the SAMI clusters, we find 29$\pm$4 per cent of SF galaxies have concentrated \update{star-formation} (defined by $C$-index < $-0.2$). This number is significantly higher than the fraction observed in the ungrouped sample (9$\pm$2 per cent). The different $C$-index distributions between clusters and lower halo-mass environments are confirmed by a K-S test; in particular, there is a tail of galaxies towards low $C$-index in clusters that is not visible in ungrouped environments. These observations match the likely effect of the  well-accepted environmental quenching mechanism in clusters, ram-pressure stripping \citep[]{Gunn1972}. This can expel the gas from the disc resulting in low or no star-formation in the outskirts of galaxies \citep[e.g.][]{Bekki_2013}. For SF-concentrated galaxies, quenching happened earlier in the outskirts than in the centres.

Further more, we compare the radial age profiles of SF-concentrated galaxies in clusters to those of ungrouped regular galaxies. The underlying assumption is that, before entering a group/cluster, the progenitors of SF-concentrated galaxies had on average the same radial age profiles as ungrouped regular galaxies. The SF-concentrated galaxies show no significant age difference compared to ungrouped regular galaxies. The inner regions of the SF-concentrated cluster galaxies are similar to that of ungrouped regular galaxies in Fig.~\ref{fig:AGEvsRe_LW_SFR}, \ref{fig:AGEvsRe_MW_SFR} (age difference is $-0.43\pm$0.77 Gyr for $Age_\mathrm{L}$, $-0.06\pm$0.65 Gyr for $Age_\mathrm{M}$). The age difference of the outer regions between the ungrouped regular galaxies and concentrated cluster galaxies is $0.19\pm$0.21 Gyr for $Age_\mathrm{L}$, 0.40$\pm$0.61 Gyr for $Age_\mathrm{M}$ (\update{1-2 $R_\mathrm{e}$})\update{, which is not significant.} The $D_\mathrm{n}$4000 index and $H\delta_\mathrm{A}$ index radial profiles also support our age measurements. In summary, SF-concentrated galaxies have lower outer \update{star-formation} and maybe undergoing `outside-in' quenching. Since the galaxies with concentrated \update{star-formation} show no significant age difference, this implies that quenching in the centre follows quickly after the outskirts are quenched.

Regular galaxies in clusters do not have centrally concentrated SF, rather they have older centres with younger outskirts. They do not show evidence of environmental quenching. The phase-space diagram (Fig.~\ref{fig:phasespace}) indicates that regular galaxies are less likely to be near the centre of the clusters. 

\begin{figure}
	\includegraphics[width=\columnwidth]{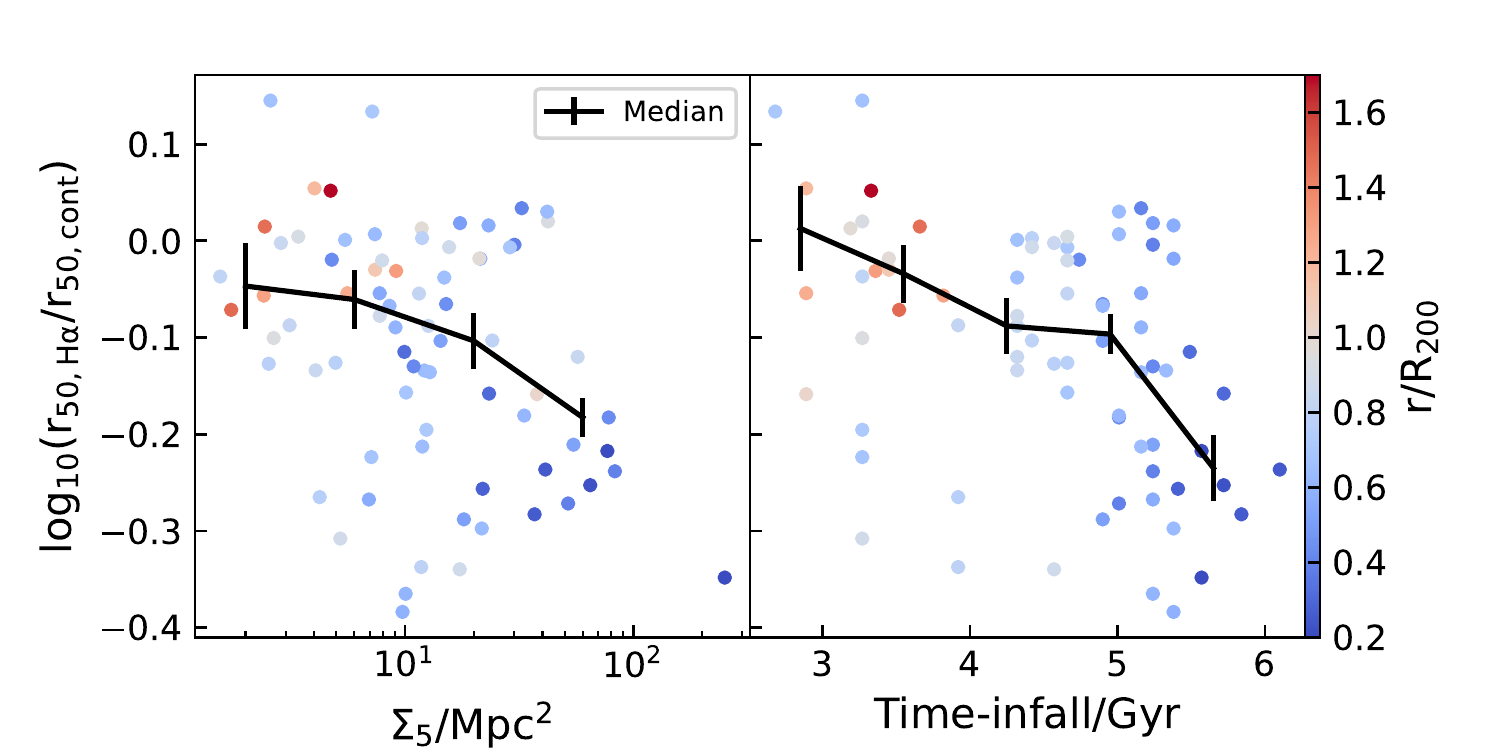}
    \caption{The left panel shows $C$-index versus $\Sigma_{5}$ for cluster galaxies colour-coded by projected distance from the centre of the cluster. The right panel shows $C$-index versus Time-infall \update{(the time since galaxies fall into clusters)} for the same galaxies. The medians are shown in black lines with uncertainties. We find the $C$-index decreases with increasing infall time-scales. }
    \label{fig:intalltime}
\end{figure}

To further investigate the role of environment on SF-concentrated galaxies, in Fig.~\ref{fig:intalltime}, we consider two related but different environment measurements: the fifth-nearest-neighbour surface density (labelled $\Sigma_{5}$; Secion.~\ref{sec:envirmetric}; left panel) and time since galaxies fall into clusters (labelled Time-infall, right panel). Time-infall has been estimated from the location of individual galaxies within the clusters' projected phase-space diagram \citep[][]{Rhee2020}. The time was obtained from numerical simulations, for which true Time-infall values are known, and defined as the time since the galaxy first crossed 1.5 virial radii. They divided the phase-space diagram in 32 bins, then assigned each galaxy the median Time-infall of the bin it belongs to. \citet[]{Rhee2020} estimated Time-infall values which are based on the full population of galaxies in a cluster (including passive galaxies). We only select SF galaxies that currently show evidence of quenching, so the true Time-infall will be lower than shown in Fig.~\ref{fig:intalltime} \update{and the trends we see should largely be considered qualitatively}. We estimate a typical uncertainty of ~2 Gyr as the 1-$\sigma$ range of the Time-infall distribution within a bin. Uncertainties on assigning time-scale to galaxies based on their position in the phase-space diagram is well discussed in \citet[]{Cortese2021}. Therefore we do not put much emphasis on the actual value of Time-infall, but instead assume that the relative difference in Time-infall between different galaxies is the most meaningful. Intuitively, $\Sigma_{5}$ is more related to local environment whereas Time-infall is more related to global environment (i.e. interactions between the galaxy and the cluster). Considering each panel separately, we find a Spearman rank correlation coefficient of $r = -0.29$ ($P$-value = 0.01) for the anticorrelation between $C$-index and $\Sigma_{5}$, and $r = -0.39$ ($P$-value = 0.0006) for the anticorrelation between $C$-index and Time-infall. Both correlations are statistically significant.

Clearly, Time-infall and $\Sigma_{5}$ are strongly correlated themselves ($r = 0.53$, $P$-value=8.6 $\times 10^{-7}$); to disentangle the relative importance of Time-infall and $\Sigma_{5}$, we use partial correlation coefficients \citep[PCC; see e.g.][]{Bait_2017, Bluck2020,Baker2022}. The PCC $r(x, z \vert y)$ measures the correlation coefficient of the two random variables $x$ and $z$ while controlling for the third variable $y$. The results show that after removing the correlation between $C$-index and time-since infall, there is no correlation between $C$-index and $\Sigma_{5}$ [$r(C-\mathrm{index}, \Sigma_5 \vert \mathrm{Time-infall}) = -0.03$, $P=0.8$]. In contrast, after removing the correlation between $C$-index and $\Sigma_{5}$, there is still some evidence (2~$\sigma$) for an anticorrelation between $C$-index and Time-infall [$r(C\mathrm{-index}, \mathrm{Time-infall} \vert \Sigma_5) = -0.28$, $P=0.02$]. This result is all the more striking if we consider that measurements of Time-infall are much more uncertain than measurements of $\Sigma_{5}$.

So the PCCs suggest that $C$-index decreases with increasing Time-infall, with no independent correlation with local environment density. This implies that variations to the $C$-index with respect to ordinary, regular galaxies are likely due to the global cluster environment, not to interactions with nearby galaxies. The large scatter in the $C$-index vs Time-infall diagram could be due to either the large measurement uncertainties on Time-infall, or, alternatively, to other physical properties other than Time-infall contributing to determine $C$-index. For example, regular galaxies with a long Time-infall could be due to quenching depending on the detailed orbit within the halo. This scenario is broadly consistent with a `delayed-then-rapid' model, when these galaxies most recent fall into a current host halo: the first 2-4 Gyr after infall being unaffected, then star-formation quenches rapidly with an e-folding time of $\le$ 0.8 Gyr \citep[][]{Wetzel_2013}. \citet[]{Wetzel_2013} argued the unaffected 2-4 Gyr is related to group preprocessing, where galaxies have not fallen into a dense enough environment like the cluster centre.

\update{\citet{Owers2019} used a small sample (25) of SF galaxies with strong Balmer absorption in the SAMI survey to study environmental quenching. They found that the majority (53 per cent) of cluster H$\delta$ strong galaxies (HDSG) show star formation in the centres, while the HDS regions in the outskirts, which is consistent with `outside-in' quenching. Along with the projected phase-space diagram, they also found that cluster HDSGs are consistent with an infalling population that has entered the central 0.5$r_{200}$ within the last $\sim$ 1 Gyr, which shows evidence of recent quenching of star formation. If the SF-concentrated galaxies in the cluster we discuss are the same as in \citet{Owers2019}, they may also have just passed within 0.5 virial radii of the cluster within the last $\sim$ 1 Gyr.} The age difference in the outer region for cluster SF-concentrated galaxies compared to regular ungrouped galaxies is 0.23$\pm$0.33 Gyr for $Age_\mathrm{L}$, suggesting that once quenching started, it happened on a short time-scale.

\subsection{Quenching in high-mass groups}

There is a slightly lower fraction (19$\pm$4 per cent) of SF-concentrated galaxies in high-mass groups compared to clusters but more than the ungrouped environment (9$\pm$2 per cent) and low-mass groups (8$\pm$3 per cent). The K-S test on the $C$-index distribution shows the $C$-index distribution is not the same as the ungrouped and low-mass group galaxies. The stellar population radial profiles also show a distinctive signature of a young centre and older outskirts for galaxies with $10^{9.5-11.5}M_{\odot}$ $M_{\star}$. The observations suggest that \update{star-formation} quenching happened in the outskirts first for SF-concentrated galaxies.

\update{Similar to} clusters, we assume that before entering a group/cluster, the average age profile is like the ungrouped galaxies. \update{This assumption means that we are looking at the integrated impact of environments, including possible pre-processing prior to entering the current group. We look into the integrated environmental quenching processes because with the SAMI data, it is impossible to rule out whether galaxies have been affected by previous haloes. However, compared with ungrouped galaxies, regular high-mass group galaxies have the same age radial profile, therefore, there is little evidence of pre-processing in high-mass groups. This is consistent with other research that shows pre-processing is mild prior to entering groups \citep[e.g.][]{Vijayaraghavan2013, Hou2014}.} \updateagain{Therefore, we do not further discuss the effect of pre-processing here.}

The age difference of the outer region between the ungrouped regular galaxies and high-mass group SF-concentrated galaxies is 1.83$\pm$0.38 Gyr for $Age_\mathrm{L}$, 1.34$\pm$0.56 Gyr for $Age_\mathrm{M}$. The great difference suggests that \update{not only has quenching happened first in the outskirts of these galaxies, but that there is also a }significant delay between the quenching of the outer disc in groups and the total quenching of a galaxy (including the central regions). We do find the $D_\mathrm{n}$4000 difference \update{at large radius between high-mass groups and ungrouped regular galaxies} which is consistent with \citet[]{Spindler2018}, who used $D_\mathrm{n}$4000 as SFR proxies in MaNGA group that showed evidence for an `outside-in' environment quenching. We also observe that the regular SF galaxies in groups, without concentrated SF, have the same age profile as ungrouped regular galaxies (Fig.~\ref{fig:AGEvsRe_LW_SFR}). Therefore, there are a large fraction of galaxies that are not yet impacted significantly by this environment.

One possible mechanism is that galaxies are partially stripped of gas by ram-pressure in high-mass groups, suppressing star-formation in the outer disc \citep[e.g.][]{schaefer2019, Vaughan2020}, while central star-formation remains. \update{Partial ram-pressure stripping is also studied by hydrodynamical simulation models, for example, \citet[]{Steinhauser2016} found that complete gas stripping in discs only happen in extreme cases and partial ram-pressure stripping can lead a star-formation enhancement in the galaxy centre. Later, \citet[]{ Stevens2017} found that stripping of hot gas from around the satellites depends on orbit and environments.} When comparing the $Age_\mathrm{L}$, which is weighted towards the most recent star-formation, the age difference of high-mass group SF-concentrated galaxies (1.83$\pm$0.38 Gyr) is larger than that of clusters ($0.19\pm$0.21 Gyr) suggesting that clusters are more efficient at removing cold gas from infalling galaxies. This is consistent with \citet{Foltz_2018}, who found quenching timescales to be faster in clusters relative to groups, suggesting that properties of the host halo are responsible for quenching high-mass galaxies. \update{For galaxies in 10$^{7.5-9.5}M_{\odot}$ $M_{\star}$ bin, we also find there are a greater fraction of SF-concentrated galaxies in high-mass groups (24$\pm$7 per cent) compared to low-mass groups (9$\pm$2 per cent) and ungrouped regions (7$\pm$2 per cent). Environmental effects on low $M_{\star}$ galaxies are also supported by simulation results \citep[e.g.][]{Cora2019}. }

An alternative mechanism is that galaxy-galaxy interactions in groups such as tidal interactions and close-pair interactions \citep[e.g.][]{Hernquist1989,For2021} could cause enhancement of central star-formation. The age difference in the inner region between high-mass group SF-concentrated galaxies and ungrouped regular galaxy is $-0.14\pm0.25$ for $Age_\mathrm{L}$ and $-0.53\pm0.57$ for $Age_\mathrm{M}$. There is no significant difference in ages, suggesting no significant recent central \update{star-formation} burst in high-mass groups for SF-concentrated galaxies. Thus, the fraction (19$\pm$4 per cent) of SF-concentrated galaxies in high-mass groups are more likely caused by quenching in the outer disc.

\subsection{Ungrouped and low-mass group environment}

Ungrouped and low-mass groups overlap in halo mass ranges ($\leq 10^{12.5} M_{\odot}$). They both have low local densities $\Sigma_{5}$ around 1 Mpc$^{-2}$. The $C$-index distributions for ungrouped and low-mass groups have a majority of galaxies with $C$-index around 0 with a small tail to low $C$-indices. The K-S test on $C$-index between ungrouped and low-mass groups is not significantly different with $p$-value=0.397. The stellar population radial profiles show that regular galaxies have an older centre with younger outskirts, while the SF-concentrated galaxies have a relatively younger centre. Regular galaxies are consistent with an `inside-out' mass buildup with stars in the centre forming early. For SF-concentrated galaxies, they have ongoing central SF. Note that this is caused by our sample selection with low $C$-indices. There is little age difference in the outer region for concentrated and regular galaxies ($-0.36\pm$0.12 Gyr for $Age_\mathrm{L}$, $-0.34\pm$0.47 Gyr for $Age_\mathrm{M}$), which shows no \update{star-formation} quenching in their outskirts which might be expected given their very low density environments. Therefore, the SF-concentrated galaxies in this case are not generally caused by quenching in the outer disc. The ungrouped regular galaxies are used as the references compared to higher halo mass environments as there is little evidence for environmental quenching in these galaxies. For low-mass group SF-concentrated galaxies, the age difference compared to ungrouped regular galaxies in the inner region is $-0.97\pm$0.23 Gyr for $Age_\mathrm{L}$, $-1.69\pm$0.65 Gyr for $Age_\mathrm{M}$. The age difference in the inner region between ungrouped SF-concentrated galaxies and ungrouped regular galaxies is $-0.89\pm$0.19 Gyr for $Age_\mathrm{L}$, $-3.70\pm$0.29 Gyr for $Age_\mathrm{M}$. The younger ages for SF-concentrated galaxies compared to regular galaxies show there is recent star-formation in the centres in ungrouped regions and low-mass groups. The low $C$-index is then likely to be driven by an enhancement of central star-formation in low-mass groups where the close-pair interactions may happen \citep[e.g.][]{Larson1978,Ellison2008,Moreno2015}.

\section{Conclusions}\label{sec:conclusion}

It is important to \update{study} the star-formation quenching mechanisms acting in different environments to \update{understand} galaxy evolution. IFS surveys have provided a crucial tool to study spatially-resolved star-formation. With the SAMI Galaxy Survey DR3, we present star-formation concentration index [$C$-index, log($r_\mathrm{50,H\alpha}/r_\mathrm{50,cont}$)] and star-formation quenching time-scale proxies in four different environments. The four halo mass intervals are ungrouped (not classified in a group galaxy in the GAMA catalogue), low-mass groups ($M_{200}$ $\leq 10^{12.5} M_{\odot}$), high-mass groups ($M_{200}$ within $10^{12.5-14} M_{\odot}$) and clusters ($M_{200}$ $> 10^{14} M_{\odot}$). $Age_\mathrm{L}$ and $Age_\mathrm{M}$ that are calculated from stellar population spectral fits, are used to estimate the quenching time-scales as well as $D_\mathrm{n}$4000 and $H\delta_\mathrm{A}$ indices. Our sample is separated into two types of galaxy by $C$-index cut: regular galaxies with $C$-index $\ge -0.2$ and SF-concentrated galaxies with $C$-index $< -0.2$. Our conclusions can be summarised as follows:

\begin{enumerate}
\item For galaxies with $M_{\star}$ between 10$^{9.5-11.5}$ $M_{\odot}$, there is a larger fraction of SF-concentrated galaxies in high-mass groups (19$\pm$4 per cent) and clusters (29$\pm$4 per cent) compared to ungrouped (9$\pm$2 per cent) and low-mass groups (8$\pm$3 per cent). The K-S test results confirm the differences between $C$-index distributions are significant in different halo masses. We also compare the fifth-nearest-neighbour surface density ($\Sigma_{5}$), and find $C$-index shows a significant difference only when $\Sigma_{5}$ $>$ 10 Mpc$^{-2}$. SF-concentrated galaxies with non-SF outskirts in high-mass groups and clusters are consistent with `outside-in' quenching. 

\item For galaxies with $M_{\star}$ between 10$^{7.5-9.5}$ $M_{\odot}$, the fraction of SF-concentrated galaxies is 7$\pm$2 per cent for ungrouped regions, 9$\pm$2 per cent  for low-mass groups and 24$\pm$7 per cent for high-mass groups (the SAMI cluster sample does not have low $M_{\star}$ galaxies). The K-S test results confirm the $C$-index distributions are significantly different between high-mass groups and ungrouped galaxies. Environments affect star-formation quenching for SF-concentrated galaxies in 10$^{7.5-9.5}$ $M_{\odot}$ mass bin as well.

\item Regular galaxy stellar population age profiles in all environments show that these galaxies have old centres with young outskirts which is consistent with `inside-out' galaxy formation \citep[e.g.][]{Perez2013, Gonzalez2015}. 

\item We assume that prior to entering a group/cluster, the average age profile is like the ungrouped galaxies. SF-concentrated galaxies in high-mass groups show older outskirts comparing to ungrouped regular galaxies. The age difference in the outer 1-2$R_\mathrm{e}$ regions between SF-concentrated galaxies in high-mass groups and regular ungrouped galaxies (1.83$\pm$0.38 Gyr for $Age_\mathrm{L}$ and 1.34$\pm$0.56 Gyr for $Age_\mathrm{M}$) suggests that star-formation quenching happened in the outskirts first for SF-concentrated galaxies and there is a significant delay between the quenching of the outer disc in groups and the total quenching of a galaxy (including the central regions). These galaxies may be influenced by partial ram-pressure stripping.

\item In the outer regions, SF-concentrated cluster galaxies and regular ungrouped galaxies have no significant age difference ($0.19\pm$0.21 Gyr for $Age_\mathrm{L}$ and 0.40$\pm$0.61 Gyr for $Age_\mathrm{M}$). This implies quenching in the centre is followed quickly after the outskirts are quenched. \update{Combining results of high-mass groups, these findings suggest a higher efficiency of ram-pressure stripping in clusters.}

\item SF-concentrated galaxies in ungrouped and low-mass groups show a similar age in the outskirts compared to regular ungrouped galaxies. This implies no early quenching in the outskirts. The ages in the inner regions for SF-concentrated galaxies in low-mass groups and ungrouped regions are smaller than the age of ungrouped regular galaxies. There is recent star-formation in SF-concentrated galaxies in ungrouped regions and low-mass groups which could be driven by close-pair interactions. 
\end{enumerate}

SF-concentrated galaxies in high-mass groups and clusters both show environmental driven `outside-in' quenching. Compared to ungrouped regular galaxies, the age difference in the outer regions of SF-concentrated galaxies is larger in high-mass groups than in clusters, which is consistent with cluster star-formation quenching being more rapid than star-formation quenching in high-mass groups \update{with more efficient ram-pressure stripping}.

\section*{Acknowledgements}
The SAMI Galaxy Survey is based on observations made at the Anglo-Australian Telescope. SAMI was developed jointly by the University of Sydney and the Australian Astronomical Observatory (AAO). The SAMI input catalogue is based on data taken from the Sloan Digital Sky Survey, the GAMA Survey and the VST ATLAS Survey. The SAMI Galaxy Survey is supported by the Australian Research Council (ARC) Centre of Excellence ASTRO 3D (CE170100013) and CAASTRO (CE110001020), and other participating institutions. The SAMI Galaxy Survey website is \url{http://sami-survey.org/}.

JJB acknowledges support of an Australian Research Council Future Fellowship (FT180100231). FDE acknowledges funding through the ERC Advanced grant 695671 ``QUENCH'' and support by the Science and Technology Facilities Council (STFC). JBH is supported by an ARC Laureate Fellowship FL140100278. The SAMI instrument was funded by Bland-Hawthorn's former Federation Fellowship FF0776384, an ARC LIEF grant LE130100198 (PI Bland-Hawthorn) and funding from the Anglo-Australian Observatory. SB acknowledges funding support from the Australian Research Council through a Future Fellowship (FT140101166) ORCID - 0000-0002-9796-1363. M.S.O. acknowledges the funding support from the Australian Research Council through a Future Fellowship (FT140100255). AMM acknowledges support from the National Science Foundation under Grant No. 2009416. JvdS acknowledges support of an Australian Research Council Discovery Early Career Research Award (project number DE200100461) funded by the Australian Government.

GAMA is a joint European-Australasian project based around a spectroscopic campaign using the Anglo-Australian Telescope. The GAMA input catalogue is based on data taken from the SDSS and the UKIRT Infrared Deep Sky Survey. Complementary imaging of the GAMA regions is being obtained by a number of independent survey programmes including GALEX MIS, VST KiDS, VISTA VIKING, WISE, Herschel-ATLAS, GMRT ASKAP providing UV to radio coverage. GAMA is funded by the STFC (UK), the ARC (Australia), the AAO, and the participating institutions. The GAMA website is \url{http://www.gama-survey.org/}.

\section*{Data Availability}

All observational data presented in this paper are available from Astronomical Optics' Data Central service at https://datacentral.org.au/ as part of the SAMI Galaxy Survey Data Release 3.



\bibliographystyle{mnras}
\bibliography{PHD_ref} 




\appendix


\bsp	
\label{lastpage}
\end{document}